\documentclass[aip,apl,amsmath,amssymb,reprint]{revtex4-1}
\usepackage{graphicx}
\usepackage{dcolumn}
\usepackage{amsmath}
\usepackage{amssymb}
\usepackage[colorlinks,urlcolor=blue,citecolor=blue,linkcolor=blue]{hyperref}
\usepackage{url}
\usepackage{natbib}
\usepackage{color}

\begin{document}

\title{Probing the features of electron dispersion by tunneling between slightly twisted bilayer graphene sheets}

\author{Alexey A. Sokolik}
\email{asokolik@hse.ru}
\affiliation{Institute for Spectroscopy, Russian Academy of Sciences, 108840 Troitsk, Moscow, Russia}
\affiliation{National Research University Higher School of Economics, 109028 Moscow, Russia}

\author{Azat F. Aminov}
\affiliation{National Research University Higher School of Economics, 109028 Moscow, Russia}
\affiliation{Institute of Microelectronics Technology and High Purity Materials, Russian Academy of Sciences, 142432 Chernogolovka, Russia}

\author{Evgenii E. Vdovin}
\author{Yurii N. Khanin}
\affiliation{Institute of Microelectronics Technology and High Purity Materials, Russian Academy of Sciences, 142432 Chernogolovka, Russia}

\author{Mikhail A. Kashchenko}
\affiliation{Programmable Functional Materials Lab, Center for Neurophysics and Neuromorphic
Technologies, 127495 Moscow, Russia}
\affiliation{Moscow Center for Advanced Studies, 123592 Moscow, Russia}

\author{Denis A. Bandurin}
\affiliation{Department of Materials Science and Engineering, National University of Singapore, 117575 Singapore, Singapore}
\affiliation{Programmable Functional Materials Lab, Center for Neurophysics and Neuromorphic
Technologies, 127495 Moscow, Russia}
\affiliation{Institute for Functional Intelligent Materials, National University of Singapore, 117544 Singapore, Singapore}

\author{Davit A. Ghazaryan}
\affiliation{Laboratory of Advanced Functional Materials, Yerevan State University, 0025 Yerevan, Armenia}
\affiliation{Institute for Functional Intelligent Materials, National University of Singapore, 117544 Singapore, Singapore}

\author{Sergey V. Morozov}
\affiliation{Institute of Microelectronics Technology and High Purity Materials, Russian Academy of Sciences, 142432 Chernogolovka, Russia}

\author{Kostya S. Novoselov}
\email{kostya@nus.edu.sg}
\affiliation{Institute for Functional Intelligent Materials, National University of Singapore, 117544 Singapore, Singapore}

\begin{abstract}
Tunneling conductance between two bilayer graphene (BLG) sheets separated by 2 nm-thick insulating barrier was measured in two devices with the twist angles between BLGs less than 1°. At small bias voltages, the tunneling occurs with conservation of energy and momentum at the points of intersection between two relatively shifted Fermi circles. Here, we experimentally found and theoretically described signatures of electron-hole asymmetric band structure of BLG: since holes are heavier, the tunneling conductance is enhanced at the hole doping due to the higher density of states. Another key feature of BLG that we explore is gap opening in a vertical electric field with a strong polarization of electron wave function at van Hove singularities near the gap edges. This polarization, by shifting electron wave function in one BLG closer to or father from the other BLG, gives rise to asymmetric tunneling resonances in the conductance around charge neutrality points, which result in strong sensitivity of the tunneling current to minor changes of the gate voltages. The observed phenomena are reproduced by our theoretical model taking into account electrostatics of the dual-gated structure, quantum capacitance effects, and self-consistent gap openings in both BLGs.
\end{abstract}

\maketitle

Recent technological advancements allowed fabrication of heterostructures with graphene or other two-dimensional conducting materials separated by ultrathin tunneling barriers \cite{Geim2013,Novoselov2016}. Easy tunability of tunneling currents in these structures by the bias and gate voltages opens the way to the creation of vertical tunneling transistors \cite{Georgiou2013,Britnell2013,Zhang2025}. High quality of graphene layers and precise alignment of their crystal lattices enables the energy- and momentum-conserving tunneling of electrons between overlapping parts of their dispersions. The tunneling current demonstrates resonant peaks and negative differential resistance, which are observed and explained in terms of electron dispersion nesting \cite{Britnell2013,Mishchenko2014,Fallahazad2015,Kim2016,Burg2017,Burg2018,Prasad2021,Ghazaryan2021,Zhang2025}.

The patterns of tunneling currents are shaped not only by the electrostatics of two-layer graphene structures and quantum capacitance effects, but also by the specifics of electronic properties of the graphene layers themselves. One example is electron chirality in single-layer graphene causing interference between tunneling current from different sublattices \cite{Greenaway2015,Wallbank2016}. Bilayer graphene is prominent for its gap opening in a vertical electric field \cite{McCann2006,Ohta2006,Min2007,Zhang2009} and sublayer polarization of van Hove singularities near its edges \cite{Yu2008,Ramasubramaniam2011,Kim2013,Joucken2021}, that also can affect the tunneling currents. Many-body effects, such as formation of interlayer excitons \cite{Li2017,Burg2018,Zeng2025} or correlated insulating phases, enrich the physics of graphene-based structures \cite{delaBarrera2022,Seiler2022}. Magnetic field adds extra dimension to the tunneling problem, by providing gauge displacement of electronic momenta at parallel orientation \cite{Wallbank2016,Burg2018,Prasad2021}, or enforcing Landau quantization if applied perpendicularly to graphene layers \cite{Fallahazad2015}.

\begin{figure}
\begin{center}
\includegraphics[width=\columnwidth]{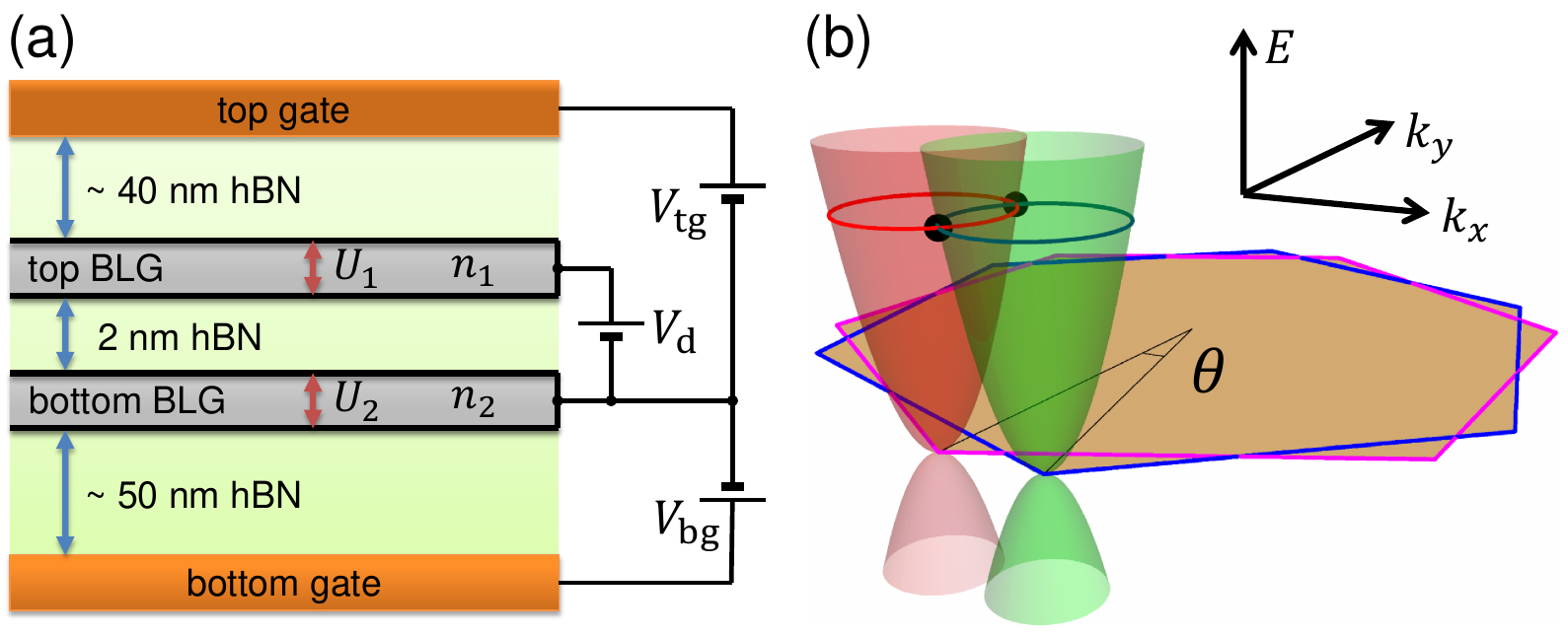}
\end{center}
\caption{\label{Fig1}(a) Schematic of a dual-gated structure consisting of top and bottom BLGs separated by the hexagonal boron nitride (hBN) barrier. (b) Electron dispersions in two BLGs twisted on the angle $\theta$ (greatly exaggerated), whose Fermi lines (red and green circles) intersect in the black points where the energy and momentum conserving tunneling occurs.}
\end{figure}

In this paper, we report the results of measurements of zero-bias electron tunneling between two BLGs in dual-gated structures (Fig.~\ref{Fig1}(a)), and our theoretical modeling, which allowed us to describe the manifestations of gap opening in BLGs and the accompanying formation of van Hove singularities in the interlayer tunneling processes, leading to significant modifications of the resonant features on the two-dimensional maps of tunneling conductance. We perform comparative study of the two structures: Device 1 with the small twist angle about $0.1^\circ$ between BLG lattices and Device 2 with the several times larger twist angle $0.73^\circ$. Preliminary experimental results on Device 2 were reported earlier \cite{Vdovin2024}. The twist gives rise to a relative shift of electron valleys in momentum space (Fig.~\ref{Fig1}(b)), and the tunneling at zero bias voltage occurs when the shifted Fermi circles of both BLGs intersect. Maps of the tunneling conductance $G$ as function of top and bottom gate voltages contain the regions where the tunneling is allowed or forbidden by energy and momentum conservation laws. Analysis of the magnitude of $G$ reveals interesting features of electron properties of BLGs: electron-hole asymmetry of carrier dispersion \cite{Mucha-Kruczynski2010,Zou2011,McCann2013} and sublayer polarization of van Hove singularities in BLG subject to a vertical electric field \cite{Yu2008,Ramasubramaniam2011,Kim2013,Joucken2021}. The observed features of the tunneling conductance are explained by the relatively simple theoretical model, which takes into account electrostatics of the dual-gated structure, quantum capacitance effects, and self-consistent gap opening in each BLG by a vertical electric field.
	
The schematic of our devices is shown in Fig.~\ref{Fig1}(a): two Bernal BLGs are separated by the 2 nm-thick hexagonal boron nitride (hBN) barrier and encapsulated by thicker (about 50 nm) top and bottom hBN layers followed by the outermost top and bottom gate electrodes. Independent contacts to each BLG allow to apply the bias $V_\mathrm{d}$ as well as top $V_\mathrm{tg}$ and bottom $V_\mathrm{bg}$ gate voltages. We perform measurements of the tunneling conductance $G$ at small bias voltage $V_\mathrm{d}\sim 1\,\mbox{mV}$, so the tunneling occurs at the common Fermi level of both BLGs. As shown in Fig.~\ref{Fig1}(b), the Fermi circles must intersect to allow energy and momentum conserving tunneling. It requires fulfillment of the triangle conditions
\begin{equation}
|k_{\mathrm{F}1}-k_{\mathrm{F}2}|\leqslant\Delta K\leqslant k_{\mathrm{F}1}+k_{\mathrm{F}2},\label{kF_ineq}
\end{equation}
where $k_{\mathrm{F}i}$ are the Fermi momenta in the top ($i=1$) and bottom ($i=2$) BLGs, $\Delta K\approx4\pi\theta/3a$ is the momentum separation of the closest valleys of top and bottom BLGs, $\theta$ is the twist angle, $a\approx2.46\,\mbox{\AA}$ is the graphene lattice constant. The Fermi momenta are related to the electron densities $n_i=\pm k_{\mathrm{F}i}^2/\pi$, which are considered as negative in the case of hole doping.

\begin{figure}
\begin{center}
\resizebox{1.0\columnwidth}{!}{\includegraphics{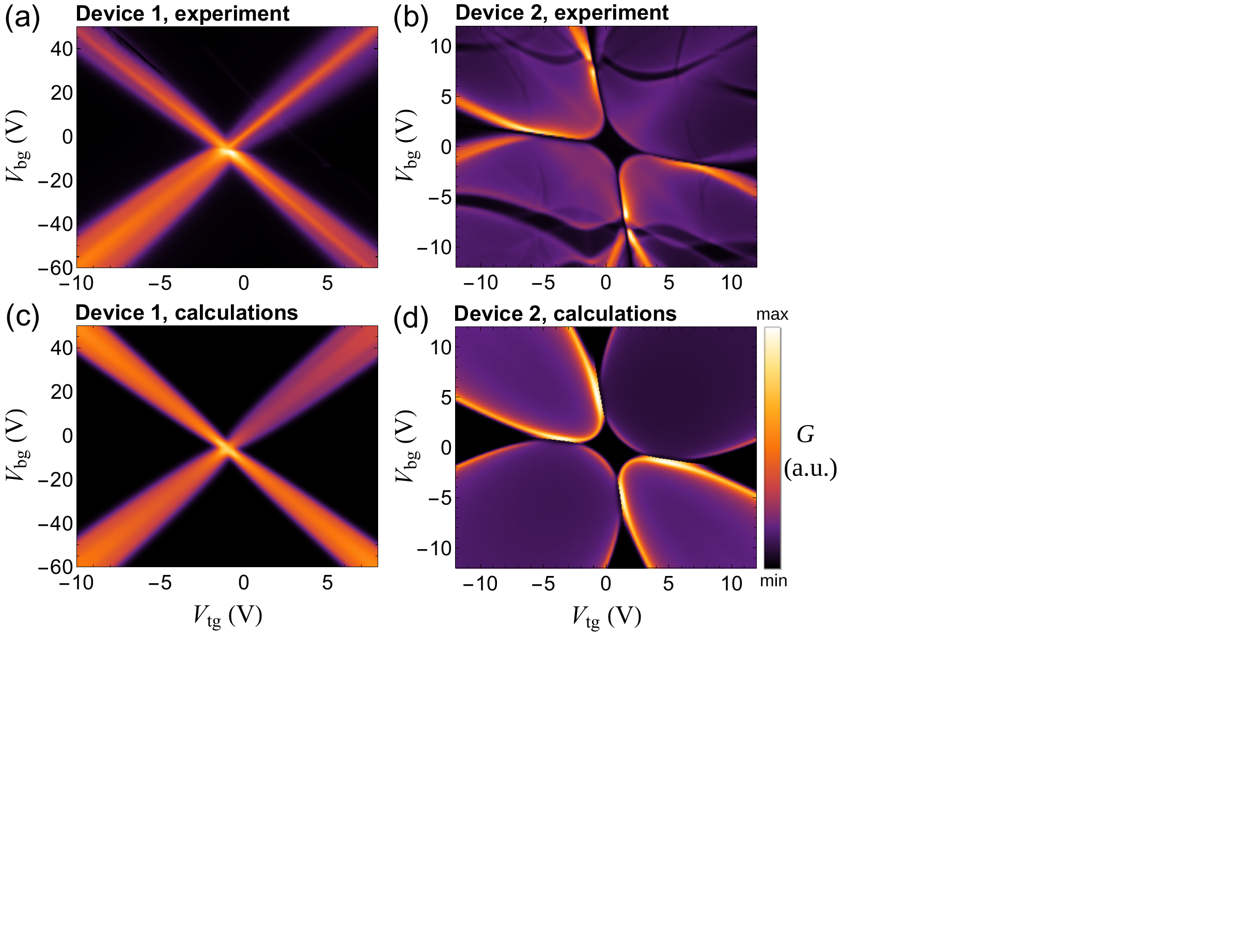}}
\end{center}
\caption{\label{Fig2}Maps of tunneling conductance (in arbitrary units) as functions of top and bottom gate voltages for (a,c) Device 1 with the combination of twist angles $\theta=0.05^\circ$ and $0.13^\circ$, and (b,d) Device 2 with the twist angle $\theta=0.73^\circ$. The upper panels (a,b) show experimental results and the lower panels (c,d) show theoretical calculations.}
\end{figure}

Fig.~\ref{Fig2}(a,b) shows the tunneling conductance in Devices 1 and 2 measured as function of top and bottom gate voltages $V_\mathrm{tg}$, $V_\mathrm{bg}$ at 4.2 K. Closeness of each gate to the corresponding BLG makes the electron density $n_1$ ($n_2$) almost proportional to $V_\mathrm{tg}$ ($V_\mathrm{bg}$). Device 1 has an additional 300 nm-thick $\mathrm{SiO}_2$ layer before the bottom gate, so much larger $V_\mathrm{bg}$ are needed to provide the same doping level (the normalized maps in terms of gate electric fields are shown in supplementary material). Quantum capacitance effects gives rise to incomplete screening of the gate electric fields, inducing the cross-talks $V_\mathrm{tg}\leftrightarrow n_2$, $V_\mathrm{bg}\leftrightarrow n_1$, and making the picture stretched along the diagonal. In Device 1 (Fig.~\ref{Fig2}(a)), the tunneling occurs close to the diagonals where the densities are equal ($n_1=n_2$) or opposite ($n_1=-n_2$), because the dispersions in two BLGs are nearly aligned. In other words, at small $\Delta K$ the inequalities (\ref{kF_ineq}) become very restrictive and boil down to $k_{\mathrm{F}1}\approx k_{\mathrm{F}2}$. In contrast, the tunneling in Device 2 (Fig.~\ref{Fig2}(a)) is allowed in four wide lobes (see the supplementary material where calculations for intermediate values of $\theta$ are shown). The tunneling is suppressed near the charge neutrality point $V_\mathrm{tg}\approx V_\mathrm{bg}\approx0$ where the carrier densities are low, and mutually shifted Fermi circles are too small to intersect. At high enough doping levels of both BLGs, the Fermi circles start to intersect. However at too large carrier density imbalance, the tunneling is suppressed again, because the smaller Fermi circle ends up inside the larger one and does not intersect it.

We describe the tunneling theoretically by calculating the current using the Fermi's golden rule
\begin{align}
I&\propto\sum_{\mathbf{k}\gamma_1\gamma_2}M_{\mathbf{k}\gamma_1\gamma_2}\int dE[n_\mathrm{F}(E)-n_\mathrm{F}(E+eV_\mathrm{d})]\nonumber\\
&\times\rho(E+\mu_1+eV_\mathrm{d}-E_{\mathbf{k}+\Delta\mathbf{K},\gamma_1})\rho(E+\mu_2-E_{\mathbf{k}\gamma_2}),
\end{align}
where integration is performed by electron energy $E$, which is measured from the Fermi level in bottom BLG (in top BLG it is lower by $eV_\mathrm{d}$). The difference of Fermi-Dirac distributions $n_\mathrm{F}(E)=[e^{E/T}+1]^{-1}$ in bottom and top BLGs is multiplied by the spectral densities $\rho(E)=(\sqrt{2\pi}\Gamma)^{-1}e^{-E^2/2\Gamma^2}$ which take into account the Gaussian broadening $\Gamma$ of electron states. The electron dispersions $E_{\mathbf{k}\gamma}$ in conduction ($\gamma=+1$) and valence ($\gamma=-1$) bands, and the chemical potentials $\mu_{1,2}$ are measured from the Dirac points of top and bottom BLGs. In the limit $V_\mathrm{d}\rightarrow0$, we obtain for the tunneling conductance $G=I/V_\mathrm{d}$:
\begin{align}
G\propto-\sum_{\mathbf{k}\gamma_1\gamma_2}M_{\mathbf{k}\gamma_1\gamma_2}&\int dE\,n'_\mathrm{F}(E)\rho(E+\mu_1-E_{\mathbf{k}+\Delta\mathbf{K},\gamma_1})\nonumber\\
&\times\rho(E+\mu_2-E_{\mathbf{k}\gamma_2}).\label{tunn_G}
\end{align}
The matrix element $M_{\mathbf{k}\gamma_1\gamma_2}$ accounts for unequal distribution of electron wave functions over BLG sublattices. Similarly to Refs.~\cite{Lane2015,Thompson2019,Prasad2021}, we assume the tunneling occurs only between the closest low-energy sublattices of BLGs: the lower sublattice A1 of top BLG and the upper sublattice B2 of bottom BLG, so the matrix element is
\begin{equation}
M_{\mathbf{k}\gamma_1\gamma_2}=|(\Psi_{\mathbf{k}\gamma_2})_\mathrm{B2}^*(\Psi_{\mathbf{k}+\Delta\mathbf{K},\gamma_1})_\mathrm{A1}|^2.\label{matr_el}
\end{equation}
The spinor wave functions $\Psi_{\mathbf{k}\gamma}$ of electron states are found by diagonalizing the effective $2\times2$  Hamiltonian, which is written in the basis \{A1, B2\} as
\begin{equation}
H=\left(\begin{array}{ll}\frac{E_+^0+E_-^0-U}2&\frac{-E_+^0+E_-^0}2\xi e^{-2i\xi\phi}\\ \frac{-E_+^0+E_-^0}2\xi e^{2i\xi\phi}&\frac{E_+^0+E_-^0+U}2\end{array}\right).\label{H_appr}
\end{equation}
Here $\xi=\pm1$ is the valley index, $\phi$ is the angle of electron momentum $\mathbf{k}$, counted from the Dirac point, with respect to the $x$ axis, $U$ is the electron energy difference between the upper (B2) and lower (A1) sublattices; $E_\pm^0=\mp\frac{\gamma_1}2\pm[(v\pm v_4)^2\hbar^2k^2+(\frac{\gamma_1}2)^2]^{1/2}$ are electron energies in conduction and valence bands at $U=0$, $v$ is the Fermi velocity, $v_4$ is its electron-hole asymmetry, and $\gamma_1$ is the vertical hopping integral in BLG. The Hamiltonian (\ref{H_appr}) provides electron energies and wave functions, which are very close (to within 10\%) to those found from the full $4\times4$ Hamiltonian of BLG \cite{McCann2013} at carrier densities $|n|\leqslant10^{13}\,\mbox{cm}^{-2}$ and the induced gaps $|U|\leqslant0.15\,\mbox{eV}$, that are well within the experimental range. At the same time, Eq.~(\ref{H_appr}) reproduces important features of electron dispersion: gap opening at $U\neq0$, parabolic dispersion at small $k$ turning into linear one at large $k$, and their electron-hole asymmetry \cite{Mucha-Kruczynski2010,Zou2011,McCann2013} (thanks to $v_4\neq0$).

To describe the electrostatics of the system, we equate the voltages $V_\mathrm{tg}$, $V_\mathrm{bg}$, $V_\mathrm{d}$ to the differences of electrochemical potentials in corresponding samples (BLGs and gates), thus taking into account quantum capacitance effects. We also take into account vertical electric fields inside each BLG, which induce the gaps $|U_i|$ in their spectra. These fields are self-consistently screened by the density response of each BLG \cite{McCann2006,Min2007,McCann2013}. The set of electrostatic equations, given in the supplementary material, are solved self-consistently for each pair $V_\mathrm{tg}$, $V_\mathrm{bg}$, and their solution is used in Eq.~(\ref{tunn_G}) to calculate the tunneling conductance.

The results of theoretical calculations of $G$ are shown in Fig.~\ref{Fig2}(c,d) for Devices 1 and 2. For Device 1, we use the combination of two twist angles $\theta=0.05^\circ$ and $\theta=0.13^\circ$ with relative contributions 1:3 to describe composition of the sample by two domains with slightly different $\theta$. Calculation parameters chosen to fit the experimental data are listed in the supplementary material. The calculations reproduce the main features of the conductance maps: existence of allowed and forbidden regions, electron-hole asymmetry between the quadrants where both BLGs are electron- ($V_\mathrm{tg,bg}>0$) and hole-doped ($V_\mathrm{tg,bg}<0$), peaks of $G$ near the boundaries of allowed regions, and asymmetry of these peaks towards the electron-hole diagonal. The dark horizontal bars in Fig.~\ref{Fig2}(b) are supposedly attributed to secondary Dirac points induced in the bottom BLG by Moire potential of hBN \cite{Yankowitz2019}. These secondary Dirac points are beyond the scope of our present study, and our model neglects the Moire coupling of BLGs with hBN, therefore the dark bars are not reproduced by the calculations.

\begin{figure}[t]
\begin{center}
\resizebox{1.0\columnwidth}{!}{\includegraphics{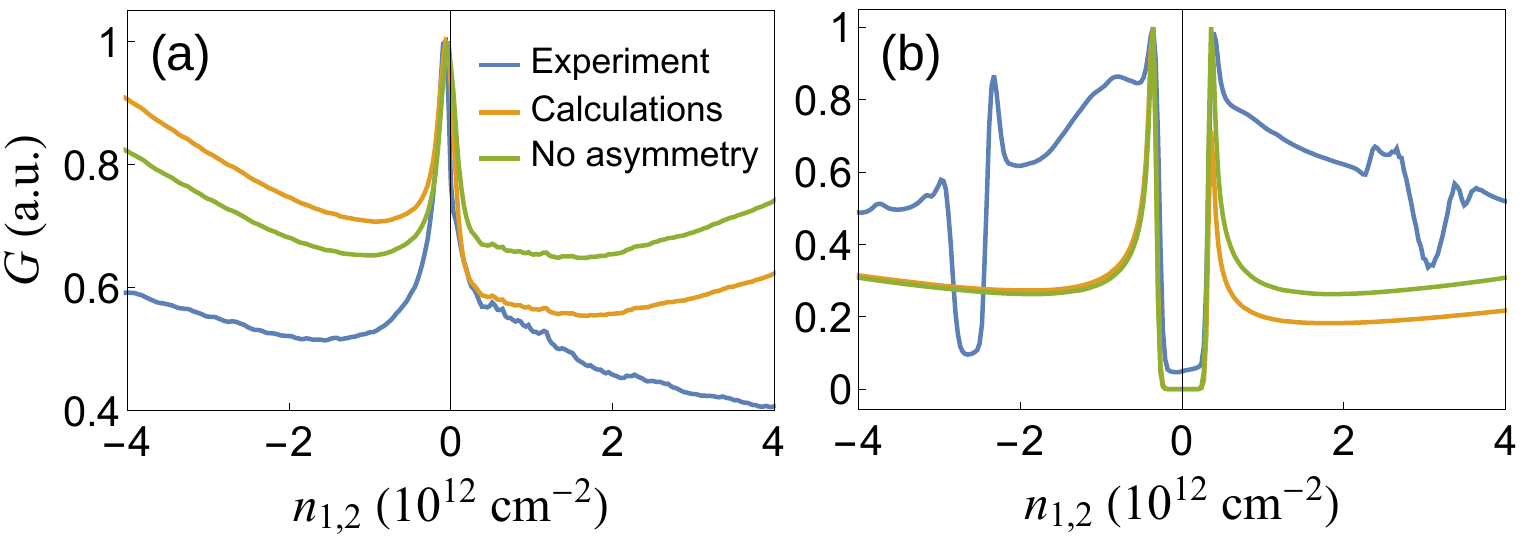}}
\end{center}
\caption{\label{Fig3}Tunneling conductance $G$ along the diagonal of equal carrier densities $n_1=n_2$ in (a) Device 1 and (b) Device 2. Theoretical calculations (orange curves) demonstrate the same hierarchy as the experiment (blue curves): $G$ is higher on the hole side ($n_{1,2}<0$) than on the electron side ($n_{1,2}>0$). In contrast, the calculations with neglecting electron-hole asymmetry of BLG spectra (green curves) provide the symmetric picture. The dips in experimental curve in (b) at $n_{1,2}\sim\pm3\times10^{12}\,\mbox{cm}^{-2}$ are caused by secondary Dirac points due to an adjacent hBN layer.}
\end{figure}

One of the nontrivial features of BLG, which manifests itself in the conductance maps, is the electron-hole asymmetry of electron dispersion caused  by the diagonal hoppings and by the on-site energy difference on different sublattices \cite{Mucha-Kruczynski2010,Zou2011,McCann2013}. Both effective mass near the Dirac point and linear asymptotics of the dispersion are asymmetric by 5-30\%. We model the asymmetry by the term $v_4=0.05v$ entering the energies $E_\pm^0$, so holes have higher cyclotron mass. As a result, the density of states (DOS) is higher in the valence band of BLG than in the conduction band. Since the tunneling conductance (\ref{tunn_G}) is approximately proportional to the product of DOS on the Fermi levels of both BLGs, $G$ should be higher in the hole-hole region $V_\mathrm{tg,bg}<0$ than in the electron-electron region $V_\mathrm{tg,bg}>0$. Fig.~\ref{Fig3}(a,b) shows the one-dimensional cross-sections of the two-dimensional tunneling conductance maps for Devices 1 and 2 along the equal-density diagonal $n_1=n_2$, whose right parts are on the electron side, and left parts are on the hole side. Both experiment and our calculations demonstrate generally higher $G$ on the hole side, in agreement with the theoretical expectations. In contrast, in the absence of electron-hole asymmetry (for $v_4=0$), the calculations result in symmetric picture.

\begin{figure}
\begin{center}
\resizebox{\columnwidth}{!}{\includegraphics{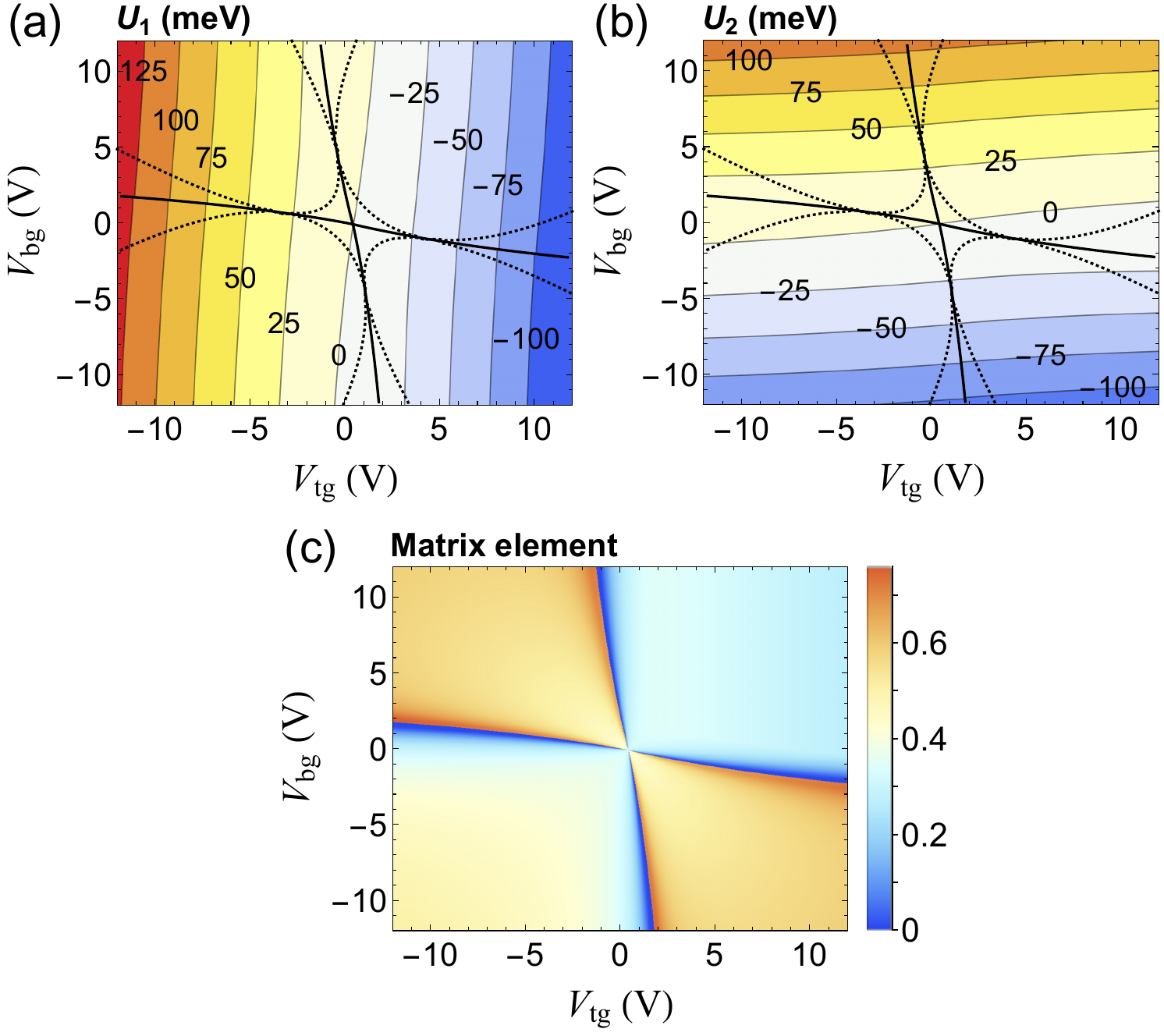}}
\end{center}
\caption{\label{Fig4}Signatures of gap opening and sublayer polarization of van Hove singularities in BLG, theoretically calculated for Device 2. (a,b) Field-induced interlayer potential differences $U_{1,2}$ in top and bottom BLGs as functions of the gate voltages (gaps are equal to their absolute values). Solid lines are charge neutrality lines $n_1=0$ or $n_2=0$, and dotted lines bound the tunneling regions. (c) Matrix element of the tunneling (\ref{matr_el}) at Fermi level.}
\end{figure}

Opening of the gaps by vertical electric fields create van Hove singularities of DOS near the gap edges \cite{McCann2006,Ohta2006,Min2007,Zhang2009}. Moreover, these singularities are strongly polarized in space \cite{Yu2008,Ramasubramaniam2011,Kim2013,Joucken2021}. For example, at $U<0$ an electron energy is lower in the upper graphene sublayer of BLG, so at the edge of valence band (where the energy is lower as well) the wave function is concentrated in the upper sublayer. Conversely, at the edge of the conduction band (at higher energy) the wave function is concentrated in the lower sublayer. Since an intensive tunneling requires concentration of the wave function in each BLG in those sublayer which is closer to the other BLG, traversal of the Fermi level through the gap reverses the polarization direction and thus drastically changes a magnitude of $G$.

As seen from Fig.~\ref{Fig4}(a,b), the calculated gaps $|U_{1,2}|$ are generally nonzero and of the same sign at charge neutrality lines (CNLs), where the Fermi level of either BLG cross the gap, so the effects of van Hove singularity polarization is most prominent. Crossing a CNL of BLG flips its Fermi level through the gap, hence changing the matrix element between large and small values (Fig.~\ref{Fig4}(c)). This behavior of the matrix element is reflected in the shape of the resonant conductance peaks, which occur near the tunneling region boundaries where the Fermi circles develop the most extended intersections at the beginning of their touching. As result, the resonant peaks of $G$ become highly asymmetric, which is indeed observed in experiment (Fig.~\ref{Fig2}(b)).

\begin{figure}
\begin{center}
\resizebox{1.0\columnwidth}{!}{\includegraphics{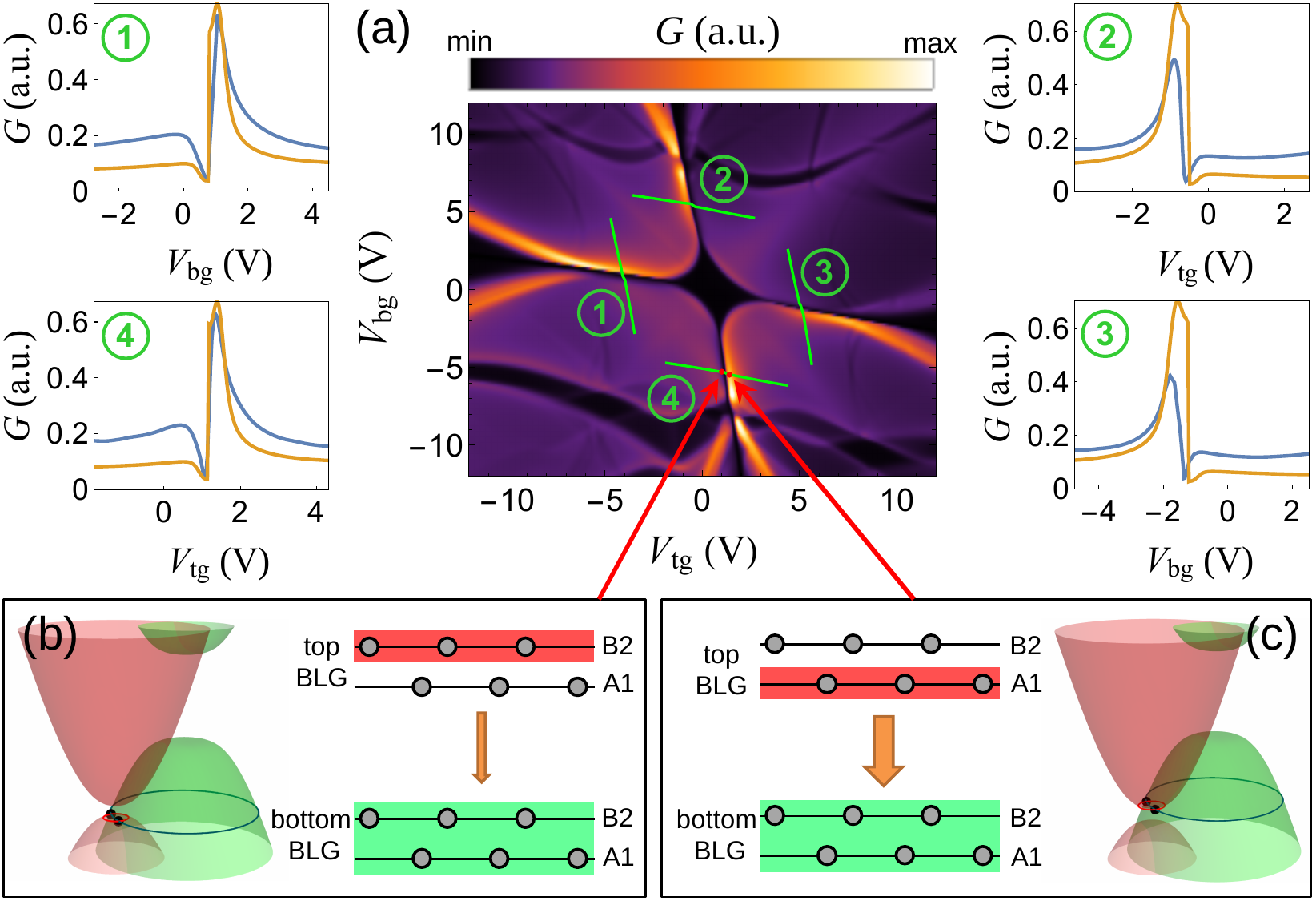}}
\end{center}
\caption{\label{Fig5}(a) Experimental map of tunneling conductance $G$ for Device 2 and its cross-sections 1-4 shown on the inset plots by blue curves. Theoretically calculated $G$ along the same cross-sections is shown by orange curves. (b) Schematic of electron localization in the upper sublayer of top BLG at the point where $G$ is suppressed by the matrix element, because the Fermi level is positioned at the ceiling of the top BLG valence band (small red circle). (c) The same, but for the point of maximum $G$ due to electron localization in the lower sublayer of top BLG, when its Fermi level is at the floor of conduction band. In both cases (b,c), the Fermi level of bottom BLG (green circle) is deep in the valence band.}
\end{figure}

Detailed forms of the asymmetric resonances are shown in Fig.~\ref{Fig5}(a) by cross-sections of the main 2D diagram along the lines where one of electron densities $n_i$ crosses zero, while the other density is constant. These cross-sections pass through touching points of the tunneling region boundaries, where the asymmetry is most noticeable. In all cross-sections we observe the tunneling resonances whose amplitudes are strikingly different on two sides of CNL, and their calculated asymmetric shapes are in good agreement with the measurements. Fig.~\ref{Fig5}(b,c) provides graphical explanation of this asymmetry for one of the cross-sections: to the left of CNL (Fig.~\ref{Fig5}(b)), the Fermi level of top BLG is located in the valence band, hence (since $U_1<0$) the wave function is concentrated in the upper sublayer, which is farther from bottom BLG, thus the tunneling is suppressed. Conversely, to the right of CNL (Fig.~\ref{Fig5}(c)) the Fermi level of top BLG is located in the conduction band, so the wave function is concentrated in the lower sublayer, which becomes closer to the bottom BLG and enhances the tunneling. Similar analysis may be carried out for other cross-sections.

In summary, the measurements of tunneling conductance between relatively twisted BLGs reveal important features of their electronic properties: electron-hole asymmetry and sublayer polarization of van Hove singularities emerging with the gap opening in vertical electric fields. The dual-gated devices allowed us to independently tune the carrier density in each BLG thus drawing gate voltage-dependent maps with the regions where the tunneling is allowed or forbidden by energy and momentum conservation laws. Here, magnitude of the tunneling conductance demonstrates the electron-hole asymmetry, and the boundaries of the corresponding tunneling regions display resonances, which are asymmetric near the charge neutrality lines in agreement with the theoretical model. Our findings highlight the great potential of quantum twisting microscopy\cite{Inbar2023} to study electron properties of 2D materials. From the applied perspective, manipulations of the tunneling conductance using the inter-sample twist, electrical doping, and gap-inducing vertical fields demonstrate high sensitivity and thus can be used to design future electronic nanoscale devices.

See the supplementary material for description of device preparation, details of electrostatic model, calculation parameters, and additional tunneling conductance maps plotted in terms of electric fields and calculated for intermediate twist angles.

\bibliography{references}

\begin{thebibliography}{2}%
\makeatletter
\providecommand \@ifxundefined [1]{%
 \@ifx{#1\undefined}
}%
\providecommand \@ifnum [1]{%
 \ifnum #1\expandafter \@firstoftwo
 \else \expandafter \@secondoftwo
 \fi
}%
\providecommand \@ifx [1]{%
 \ifx #1\expandafter \@firstoftwo
 \else \expandafter \@secondoftwo
 \fi
}%
\providecommand \natexlab [1]{#1}%
\providecommand \enquote  [1]{``#1''}%
\providecommand \bibnamefont  [1]{#1}%
\providecommand \bibfnamefont [1]{#1}%
\providecommand \citenamefont [1]{#1}%
\providecommand \href@noop [0]{\@secondoftwo}%
\providecommand \href [0]{\begingroup \@sanitize@url \@href}%
\providecommand \@href[1]{\@@startlink{#1}\@@href}%
\providecommand \@@href[1]{\endgroup#1\@@endlink}%
\providecommand \@sanitize@url [0]{\catcode `\\12\catcode `\$12\catcode
  `\&12\catcode `\#12\catcode `\^12\catcode `\_12\catcode `\%12\relax}%
\providecommand \@@startlink[1]{}%
\providecommand \@@endlink[0]{}%
\providecommand \url  [0]{\begingroup\@sanitize@url \@url }%
\providecommand \@url [1]{\endgroup\@href {#1}{\urlprefix }}%
\providecommand \urlprefix  [0]{URL }%
\providecommand \Eprint [0]{\href }%
\providecommand \doibase [0]{https://doi.org/}%
\providecommand \selectlanguage [0]{\@gobble}%
\providecommand \bibinfo  [0]{\@secondoftwo}%
\providecommand \bibfield  [0]{\@secondoftwo}%
\providecommand \translation [1]{[#1]}%
\providecommand \BibitemOpen [0]{}%
\providecommand \bibitemStop [0]{}%
\providecommand \bibitemNoStop [0]{.\EOS\space}%
\providecommand \EOS [0]{\spacefactor3000\relax}%
\providecommand \BibitemShut  [1]{\csname bibitem#1\endcsname}%
\let\auto@bib@innerbib\@empty
\bibitem [{\citenamefont {McCann}\ and\ \citenamefont
  {Koshino}(2013)}]{McCann2013}%
  \BibitemOpen
  \bibfield  {author} {\bibinfo {author} {\bibfnamefont {E.}~\bibnamefont
  {McCann}}\ and\ \bibinfo {author} {\bibfnamefont {M.}~\bibnamefont
  {Koshino}},\ }\bibfield  {title} {\bibinfo {title} {The electronic properties
  of bilayer graphene},\ }\href {https://doi.org/10.1088/0034-4885/76/5/056503}
  {\bibfield  {journal} {\bibinfo  {journal} {Reports on Progress in Physics}\
  }\textbf {\bibinfo {volume} {76}},\ \bibinfo {pages} {056503} (\bibinfo
  {year} {2013})}\BibitemShut {NoStop}%
\bibitem [{\citenamefont {Laturia}\ \emph {et~al.}(2018)\citenamefont
  {Laturia}, \citenamefont {Van~de Put},\ and\ \citenamefont
  {Vandenberghe}}]{Laturia2018}%
  \BibitemOpen
  \bibfield  {author} {\bibinfo {author} {\bibfnamefont {A.}~\bibnamefont
  {Laturia}}, \bibinfo {author} {\bibfnamefont {M.~L.}\ \bibnamefont {Van~de
  Put}},\ and\ \bibinfo {author} {\bibfnamefont {W.~G.}\ \bibnamefont
  {Vandenberghe}},\ }\bibfield  {title} {\bibinfo {title} {Dielectric
  properties of hexagonal boron nitride and transition metal dichalcogenides:
  from monolayer to bulk},\ }\href {https://doi.org/10.1038/s41699-018-0050-x}
  {\bibfield  {journal} {\bibinfo  {journal} {npj 2D Materials and
  Applications}\ }\textbf {\bibinfo {volume} {2}},\ \bibinfo {pages} {6}
  (\bibinfo {year} {2018})}\BibitemShut {NoStop}%
\end{thebibliography}%


\begin{thebibliography}{34}%
\makeatletter
\providecommand \@ifxundefined [1]{%
 \@ifx{#1\undefined}
}%
\providecommand \@ifnum [1]{%
 \ifnum #1\expandafter \@firstoftwo
 \else \expandafter \@secondoftwo
 \fi
}%
\providecommand \@ifx [1]{%
 \ifx #1\expandafter \@firstoftwo
 \else \expandafter \@secondoftwo
 \fi
}%
\providecommand \natexlab [1]{#1}%
\providecommand \enquote  [1]{``#1''}%
\providecommand \bibnamefont  [1]{#1}%
\providecommand \bibfnamefont [1]{#1}%
\providecommand \citenamefont [1]{#1}%
\providecommand \href@noop [0]{\@secondoftwo}%
\providecommand \href [0]{\begingroup \@sanitize@url \@href}%
\providecommand \@href[1]{\@@startlink{#1}\@@href}%
\providecommand \@@href[1]{\endgroup#1\@@endlink}%
\providecommand \@sanitize@url [0]{\catcode `\\12\catcode `\$12\catcode
  `\&12\catcode `\#12\catcode `\^12\catcode `\_12\catcode `\%12\relax}%
\providecommand \@@startlink[1]{}%
\providecommand \@@endlink[0]{}%
\providecommand \url  [0]{\begingroup\@sanitize@url \@url }%
\providecommand \@url [1]{\endgroup\@href {#1}{\urlprefix }}%
\providecommand \urlprefix  [0]{URL }%
\providecommand \Eprint [0]{\href }%
\providecommand \doibase [0]{http://dx.doi.org/}%
\providecommand \selectlanguage [0]{\@gobble}%
\providecommand \bibinfo  [0]{\@secondoftwo}%
\providecommand \bibfield  [0]{\@secondoftwo}%
\providecommand \translation [1]{[#1]}%
\providecommand \BibitemOpen [0]{}%
\providecommand \bibitemStop [0]{}%
\providecommand \bibitemNoStop [0]{.\EOS\space}%
\providecommand \EOS [0]{\spacefactor3000\relax}%
\providecommand \BibitemShut  [1]{\csname bibitem#1\endcsname}%
\let\auto@bib@innerbib\@empty
\bibitem [{\citenamefont {Geim}\ and\ \citenamefont
  {Grigorieva}(2013)}]{Geim2013}%
  \BibitemOpen
  \bibfield  {author} {\bibinfo {author} {\bibfnamefont {A.~K.}\ \bibnamefont
  {Geim}}\ and\ \bibinfo {author} {\bibfnamefont {I.~V.}\ \bibnamefont
  {Grigorieva}},\ }\href {\doibase 10.1038/nature12385} {\bibfield  {journal}
  {\bibinfo  {journal} {Nature}\ }\textbf {\bibinfo {volume} {499}},\ \bibinfo
  {pages} {419} (\bibinfo {year} {2013})}\BibitemShut {NoStop}%
\bibitem [{\citenamefont {Novoselov}\ \emph {et~al.}(2016)\citenamefont
  {Novoselov}, \citenamefont {Mishchenko}, \citenamefont {Carvalho},\ and\
  \citenamefont {Castro~Neto}}]{Novoselov2016}%
  \BibitemOpen
  \bibfield  {author} {\bibinfo {author} {\bibfnamefont {K.~S.}\ \bibnamefont
  {Novoselov}}, \bibinfo {author} {\bibfnamefont {A.}~\bibnamefont
  {Mishchenko}}, \bibinfo {author} {\bibfnamefont {A.}~\bibnamefont
  {Carvalho}}, \ and\ \bibinfo {author} {\bibfnamefont {A.~H.}\ \bibnamefont
  {Castro~Neto}},\ }\href {\doibase 10.1126/science.aac9439} {\bibfield
  {journal} {\bibinfo  {journal} {Science}\ }\textbf {\bibinfo {volume}
  {353}},\ \bibinfo {pages} {aac9439} (\bibinfo {year} {2016})}\BibitemShut
  {NoStop}%
\bibitem [{\citenamefont {Georgiou}\ \emph {et~al.}(2013)\citenamefont
  {Georgiou}, \citenamefont {Jalil}, \citenamefont {Belle}, \citenamefont
  {Britnell}, \citenamefont {Gorbachev}, \citenamefont {Morozov}, \citenamefont
  {Kim}, \citenamefont {Gholinia}, \citenamefont {Haigh}, \citenamefont
  {Makarovsky}, \citenamefont {Eaves}, \citenamefont {Ponomarenko},
  \citenamefont {Geim}, \citenamefont {Novoselov},\ and\ \citenamefont
  {Mishchenko}}]{Georgiou2013}%
  \BibitemOpen
  \bibfield  {author} {\bibinfo {author} {\bibfnamefont {T.}~\bibnamefont
  {Georgiou}}, \bibinfo {author} {\bibfnamefont {R.}~\bibnamefont {Jalil}},
  \bibinfo {author} {\bibfnamefont {B.~D.}\ \bibnamefont {Belle}}, \bibinfo
  {author} {\bibfnamefont {L.}~\bibnamefont {Britnell}}, \bibinfo {author}
  {\bibfnamefont {R.~V.}\ \bibnamefont {Gorbachev}}, \bibinfo {author}
  {\bibfnamefont {S.~V.}\ \bibnamefont {Morozov}}, \bibinfo {author}
  {\bibfnamefont {Y.-J.}\ \bibnamefont {Kim}}, \bibinfo {author} {\bibfnamefont
  {A.}~\bibnamefont {Gholinia}}, \bibinfo {author} {\bibfnamefont {S.~J.}\
  \bibnamefont {Haigh}}, \bibinfo {author} {\bibfnamefont {O.}~\bibnamefont
  {Makarovsky}}, \bibinfo {author} {\bibfnamefont {L.}~\bibnamefont {Eaves}},
  \bibinfo {author} {\bibfnamefont {L.~A.}\ \bibnamefont {Ponomarenko}},
  \bibinfo {author} {\bibfnamefont {A.~K.}\ \bibnamefont {Geim}}, \bibinfo
  {author} {\bibfnamefont {K.~S.}\ \bibnamefont {Novoselov}}, \ and\ \bibinfo
  {author} {\bibfnamefont {A.}~\bibnamefont {Mishchenko}},\ }\href {\doibase
  10.1038/nnano.2012.224} {\bibfield  {journal} {\bibinfo  {journal} {Nature
  Nanotechnology}\ }\textbf {\bibinfo {volume} {8}},\ \bibinfo {pages} {100}
  (\bibinfo {year} {2013})}\BibitemShut {NoStop}%
\bibitem [{\citenamefont {Britnell}\ \emph {et~al.}(2013)\citenamefont
  {Britnell}, \citenamefont {Gorbachev}, \citenamefont {Geim}, \citenamefont
  {Ponomarenko}, \citenamefont {Mishchenko}, \citenamefont {Greenaway},
  \citenamefont {Fromhold}, \citenamefont {Novoselov},\ and\ \citenamefont
  {Eaves}}]{Britnell2013}%
  \BibitemOpen
  \bibfield  {author} {\bibinfo {author} {\bibfnamefont {L.}~\bibnamefont
  {Britnell}}, \bibinfo {author} {\bibfnamefont {R.~V.}\ \bibnamefont
  {Gorbachev}}, \bibinfo {author} {\bibfnamefont {A.~K.}\ \bibnamefont {Geim}},
  \bibinfo {author} {\bibfnamefont {L.~A.}\ \bibnamefont {Ponomarenko}},
  \bibinfo {author} {\bibfnamefont {A.}~\bibnamefont {Mishchenko}}, \bibinfo
  {author} {\bibfnamefont {M.~T.}\ \bibnamefont {Greenaway}}, \bibinfo {author}
  {\bibfnamefont {T.~M.}\ \bibnamefont {Fromhold}}, \bibinfo {author}
  {\bibfnamefont {K.~S.}\ \bibnamefont {Novoselov}}, \ and\ \bibinfo {author}
  {\bibfnamefont {L.}~\bibnamefont {Eaves}},\ }\href {\doibase
  10.1038/ncomms2817} {\bibfield  {journal} {\bibinfo  {journal} {Nature
  Communications}\ }\textbf {\bibinfo {volume} {4}},\ \bibinfo {pages} {1794}
  (\bibinfo {year} {2013})}\BibitemShut {NoStop}%
\bibitem [{\citenamefont {Zhang}\ \emph {et~al.}(2025)\citenamefont {Zhang},
  \citenamefont {Zhang}, \citenamefont {Zhang}, \citenamefont {Wang},
  \citenamefont {Hays}, \citenamefont {Tongay}, \citenamefont {Wang},
  \citenamefont {Han}, \citenamefont {Li}, \citenamefont {Zhang},\ and\
  \citenamefont {Song}}]{Zhang2025}%
  \BibitemOpen
  \bibfield  {author} {\bibinfo {author} {\bibfnamefont {Z.}~\bibnamefont
  {Zhang}}, \bibinfo {author} {\bibfnamefont {B.}~\bibnamefont {Zhang}},
  \bibinfo {author} {\bibfnamefont {Y.}~\bibnamefont {Zhang}}, \bibinfo
  {author} {\bibfnamefont {Y.}~\bibnamefont {Wang}}, \bibinfo {author}
  {\bibfnamefont {P.}~\bibnamefont {Hays}}, \bibinfo {author} {\bibfnamefont
  {S.~A.}\ \bibnamefont {Tongay}}, \bibinfo {author} {\bibfnamefont
  {M.}~\bibnamefont {Wang}}, \bibinfo {author} {\bibfnamefont {H.}~\bibnamefont
  {Han}}, \bibinfo {author} {\bibfnamefont {H.}~\bibnamefont {Li}}, \bibinfo
  {author} {\bibfnamefont {J.}~\bibnamefont {Zhang}}, \ and\ \bibinfo {author}
  {\bibfnamefont {A.}~\bibnamefont {Song}},\ }\href {\doibase
  10.1038/s41467-025-58720-7} {\bibfield  {journal} {\bibinfo  {journal}
  {Nature Communications}\ }\textbf {\bibinfo {volume} {16}},\ \bibinfo {pages}
  {4805} (\bibinfo {year} {2025})}\BibitemShut {NoStop}%
\bibitem [{\citenamefont {Mishchenko}\ \emph {et~al.}(2014)\citenamefont
  {Mishchenko}, \citenamefont {Tu}, \citenamefont {Cao}, \citenamefont
  {Gorbachev}, \citenamefont {Wallbank}, \citenamefont {Greenaway},
  \citenamefont {Morozov}, \citenamefont {Morozov}, \citenamefont {Zhu},
  \citenamefont {Wong}, \citenamefont {Withers}, \citenamefont {Woods},
  \citenamefont {Kim}, \citenamefont {Watanabe}, \citenamefont {Taniguchi},
  \citenamefont {Vdovin}, \citenamefont {Makarovsky}, \citenamefont {Fromhold},
  \citenamefont {Fal'ko}, \citenamefont {Geim}, \citenamefont {Eaves},\ and\
  \citenamefont {Novoselov}}]{Mishchenko2014}%
  \BibitemOpen
  \bibfield  {author} {\bibinfo {author} {\bibfnamefont {A.}~\bibnamefont
  {Mishchenko}}, \bibinfo {author} {\bibfnamefont {J.~S.}\ \bibnamefont {Tu}},
  \bibinfo {author} {\bibfnamefont {Y.}~\bibnamefont {Cao}}, \bibinfo {author}
  {\bibfnamefont {R.~V.}\ \bibnamefont {Gorbachev}}, \bibinfo {author}
  {\bibfnamefont {J.~R.}\ \bibnamefont {Wallbank}}, \bibinfo {author}
  {\bibfnamefont {M.~T.}\ \bibnamefont {Greenaway}}, \bibinfo {author}
  {\bibfnamefont {V.~E.}\ \bibnamefont {Morozov}}, \bibinfo {author}
  {\bibfnamefont {S.~V.}\ \bibnamefont {Morozov}}, \bibinfo {author}
  {\bibfnamefont {M.~J.}\ \bibnamefont {Zhu}}, \bibinfo {author} {\bibfnamefont
  {S.~L.}\ \bibnamefont {Wong}}, \bibinfo {author} {\bibfnamefont
  {F.}~\bibnamefont {Withers}}, \bibinfo {author} {\bibfnamefont {C.~R.}\
  \bibnamefont {Woods}}, \bibinfo {author} {\bibfnamefont {Y.-J.}\ \bibnamefont
  {Kim}}, \bibinfo {author} {\bibfnamefont {K.}~\bibnamefont {Watanabe}},
  \bibinfo {author} {\bibfnamefont {T.}~\bibnamefont {Taniguchi}}, \bibinfo
  {author} {\bibfnamefont {E.~E.}\ \bibnamefont {Vdovin}}, \bibinfo {author}
  {\bibfnamefont {O.}~\bibnamefont {Makarovsky}}, \bibinfo {author}
  {\bibfnamefont {T.~M.}\ \bibnamefont {Fromhold}}, \bibinfo {author}
  {\bibfnamefont {V.~I.}\ \bibnamefont {Fal'ko}}, \bibinfo {author}
  {\bibfnamefont {A.~K.}\ \bibnamefont {Geim}}, \bibinfo {author}
  {\bibfnamefont {L.}~\bibnamefont {Eaves}}, \ and\ \bibinfo {author}
  {\bibfnamefont {K.~S.}\ \bibnamefont {Novoselov}},\ }\href {\doibase
  10.1038/nnano.2014.187} {\bibfield  {journal} {\bibinfo  {journal} {Nature
  Nanotechnology}\ }\textbf {\bibinfo {volume} {9}},\ \bibinfo {pages} {808}
  (\bibinfo {year} {2014})}\BibitemShut {NoStop}%
\bibitem [{\citenamefont {Fallahazad}\ \emph {et~al.}(2015)\citenamefont
  {Fallahazad}, \citenamefont {Lee}, \citenamefont {Kang}, \citenamefont {Xue},
  \citenamefont {Larentis}, \citenamefont {Corbet}, \citenamefont {Kim},
  \citenamefont {Movva}, \citenamefont {Taniguchi}, \citenamefont {Watanabe},
  \citenamefont {Register}, \citenamefont {Banerjee},\ and\ \citenamefont
  {Tutuc}}]{Fallahazad2015}%
  \BibitemOpen
  \bibfield  {author} {\bibinfo {author} {\bibfnamefont {B.}~\bibnamefont
  {Fallahazad}}, \bibinfo {author} {\bibfnamefont {K.}~\bibnamefont {Lee}},
  \bibinfo {author} {\bibfnamefont {S.}~\bibnamefont {Kang}}, \bibinfo {author}
  {\bibfnamefont {J.}~\bibnamefont {Xue}}, \bibinfo {author} {\bibfnamefont
  {S.}~\bibnamefont {Larentis}}, \bibinfo {author} {\bibfnamefont
  {C.}~\bibnamefont {Corbet}}, \bibinfo {author} {\bibfnamefont
  {K.}~\bibnamefont {Kim}}, \bibinfo {author} {\bibfnamefont {H.~C.~P.}\
  \bibnamefont {Movva}}, \bibinfo {author} {\bibfnamefont {T.}~\bibnamefont
  {Taniguchi}}, \bibinfo {author} {\bibfnamefont {K.}~\bibnamefont {Watanabe}},
  \bibinfo {author} {\bibfnamefont {L.~F.}\ \bibnamefont {Register}}, \bibinfo
  {author} {\bibfnamefont {S.~K.}\ \bibnamefont {Banerjee}}, \ and\ \bibinfo
  {author} {\bibfnamefont {E.}~\bibnamefont {Tutuc}},\ }\href {\doibase
  10.1021/nl503756y} {\bibfield  {journal} {\bibinfo  {journal} {Nano Letters}\
  }\textbf {\bibinfo {volume} {15}},\ \bibinfo {pages} {428} (\bibinfo {year}
  {2015})}\BibitemShut {NoStop}%
\bibitem [{\citenamefont {Kim}\ \emph {et~al.}(2016)\citenamefont {Kim},
  \citenamefont {Yankowitz}, \citenamefont {Fallahazad}, \citenamefont {Kang},
  \citenamefont {Movva}, \citenamefont {Huang}, \citenamefont {Larentis},
  \citenamefont {Corbet}, \citenamefont {Taniguchi}, \citenamefont {Watanabe},
  \citenamefont {Banerjee}, \citenamefont {LeRoy},\ and\ \citenamefont
  {Tutuc}}]{Kim2016}%
  \BibitemOpen
  \bibfield  {author} {\bibinfo {author} {\bibfnamefont {K.}~\bibnamefont
  {Kim}}, \bibinfo {author} {\bibfnamefont {M.}~\bibnamefont {Yankowitz}},
  \bibinfo {author} {\bibfnamefont {B.}~\bibnamefont {Fallahazad}}, \bibinfo
  {author} {\bibfnamefont {S.}~\bibnamefont {Kang}}, \bibinfo {author}
  {\bibfnamefont {H.~C.~P.}\ \bibnamefont {Movva}}, \bibinfo {author}
  {\bibfnamefont {S.}~\bibnamefont {Huang}}, \bibinfo {author} {\bibfnamefont
  {S.}~\bibnamefont {Larentis}}, \bibinfo {author} {\bibfnamefont {C.~M.}\
  \bibnamefont {Corbet}}, \bibinfo {author} {\bibfnamefont {T.}~\bibnamefont
  {Taniguchi}}, \bibinfo {author} {\bibfnamefont {K.}~\bibnamefont {Watanabe}},
  \bibinfo {author} {\bibfnamefont {S.~K.}\ \bibnamefont {Banerjee}}, \bibinfo
  {author} {\bibfnamefont {B.~J.}\ \bibnamefont {LeRoy}}, \ and\ \bibinfo
  {author} {\bibfnamefont {E.}~\bibnamefont {Tutuc}},\ }\href {\doibase
  10.1021/acs.nanolett.5b05263} {\bibfield  {journal} {\bibinfo  {journal}
  {Nano Letters}\ }\textbf {\bibinfo {volume} {16}},\ \bibinfo {pages} {1989}
  (\bibinfo {year} {2016})}\BibitemShut {NoStop}%
\bibitem [{\citenamefont {Burg}\ \emph {et~al.}(2017)\citenamefont {Burg},
  \citenamefont {Prasad}, \citenamefont {Fallahazad}, \citenamefont {Valsaraj},
  \citenamefont {Kim}, \citenamefont {Taniguchi}, \citenamefont {Watanabe},
  \citenamefont {Wang}, \citenamefont {Kim}, \citenamefont {Register},\ and\
  \citenamefont {Tutuc}}]{Burg2017}%
  \BibitemOpen
  \bibfield  {author} {\bibinfo {author} {\bibfnamefont {G.~W.}\ \bibnamefont
  {Burg}}, \bibinfo {author} {\bibfnamefont {N.}~\bibnamefont {Prasad}},
  \bibinfo {author} {\bibfnamefont {B.}~\bibnamefont {Fallahazad}}, \bibinfo
  {author} {\bibfnamefont {A.}~\bibnamefont {Valsaraj}}, \bibinfo {author}
  {\bibfnamefont {K.}~\bibnamefont {Kim}}, \bibinfo {author} {\bibfnamefont
  {T.}~\bibnamefont {Taniguchi}}, \bibinfo {author} {\bibfnamefont
  {K.}~\bibnamefont {Watanabe}}, \bibinfo {author} {\bibfnamefont
  {Q.}~\bibnamefont {Wang}}, \bibinfo {author} {\bibfnamefont {M.~J.}\
  \bibnamefont {Kim}}, \bibinfo {author} {\bibfnamefont {L.~F.}\ \bibnamefont
  {Register}}, \ and\ \bibinfo {author} {\bibfnamefont {E.}~\bibnamefont
  {Tutuc}},\ }\href {\doibase 10.1021/acs.nanolett.7b01505} {\bibfield
  {journal} {\bibinfo  {journal} {Nano Letters}\ }\textbf {\bibinfo {volume}
  {17}},\ \bibinfo {pages} {3919} (\bibinfo {year} {2017})}\BibitemShut
  {NoStop}%
\bibitem [{\citenamefont {Burg}\ \emph {et~al.}(2018)\citenamefont {Burg},
  \citenamefont {Prasad}, \citenamefont {Kim}, \citenamefont {Taniguchi},
  \citenamefont {Watanabe}, \citenamefont {MacDonald}, \citenamefont
  {Register},\ and\ \citenamefont {Tutuc}}]{Burg2018}%
  \BibitemOpen
  \bibfield  {author} {\bibinfo {author} {\bibfnamefont {G.~W.}\ \bibnamefont
  {Burg}}, \bibinfo {author} {\bibfnamefont {N.}~\bibnamefont {Prasad}},
  \bibinfo {author} {\bibfnamefont {K.}~\bibnamefont {Kim}}, \bibinfo {author}
  {\bibfnamefont {T.}~\bibnamefont {Taniguchi}}, \bibinfo {author}
  {\bibfnamefont {K.}~\bibnamefont {Watanabe}}, \bibinfo {author}
  {\bibfnamefont {A.~H.}\ \bibnamefont {MacDonald}}, \bibinfo {author}
  {\bibfnamefont {L.~F.}\ \bibnamefont {Register}}, \ and\ \bibinfo {author}
  {\bibfnamefont {E.}~\bibnamefont {Tutuc}},\ }\href {\doibase
  10.1103/PhysRevLett.120.177702} {\bibfield  {journal} {\bibinfo  {journal}
  {Phys. Rev. Lett.}\ }\textbf {\bibinfo {volume} {120}},\ \bibinfo {pages}
  {177702} (\bibinfo {year} {2018})}\BibitemShut {NoStop}%
\bibitem [{\citenamefont {Prasad}\ \emph {et~al.}(2021)\citenamefont {Prasad},
  \citenamefont {Burg}, \citenamefont {Watanabe}, \citenamefont {Taniguchi},
  \citenamefont {Register},\ and\ \citenamefont {Tutuc}}]{Prasad2021}%
  \BibitemOpen
  \bibfield  {author} {\bibinfo {author} {\bibfnamefont {N.}~\bibnamefont
  {Prasad}}, \bibinfo {author} {\bibfnamefont {G.~W.}\ \bibnamefont {Burg}},
  \bibinfo {author} {\bibfnamefont {K.}~\bibnamefont {Watanabe}}, \bibinfo
  {author} {\bibfnamefont {T.}~\bibnamefont {Taniguchi}}, \bibinfo {author}
  {\bibfnamefont {L.~F.}\ \bibnamefont {Register}}, \ and\ \bibinfo {author}
  {\bibfnamefont {E.}~\bibnamefont {Tutuc}},\ }\href {\doibase
  10.1103/PhysRevLett.127.117701} {\bibfield  {journal} {\bibinfo  {journal}
  {Phys. Rev. Lett.}\ }\textbf {\bibinfo {volume} {127}},\ \bibinfo {pages}
  {117701} (\bibinfo {year} {2021})}\BibitemShut {NoStop}%
\bibitem [{\citenamefont {Ghazaryan}\ \emph {et~al.}(2021)\citenamefont
  {Ghazaryan}, \citenamefont {Misra}, \citenamefont {Vdovin}, \citenamefont
  {Watanabe}, \citenamefont {Taniguchi}, \citenamefont {Morozov}, \citenamefont
  {Mishchenko},\ and\ \citenamefont {Novoselov}}]{Ghazaryan2021}%
  \BibitemOpen
  \bibfield  {author} {\bibinfo {author} {\bibfnamefont {D.~A.}\ \bibnamefont
  {Ghazaryan}}, \bibinfo {author} {\bibfnamefont {A.}~\bibnamefont {Misra}},
  \bibinfo {author} {\bibfnamefont {E.~E.}\ \bibnamefont {Vdovin}}, \bibinfo
  {author} {\bibfnamefont {K.}~\bibnamefont {Watanabe}}, \bibinfo {author}
  {\bibfnamefont {T.}~\bibnamefont {Taniguchi}}, \bibinfo {author}
  {\bibfnamefont {S.~V.}\ \bibnamefont {Morozov}}, \bibinfo {author}
  {\bibfnamefont {A.}~\bibnamefont {Mishchenko}}, \ and\ \bibinfo {author}
  {\bibfnamefont {K.~S.}\ \bibnamefont {Novoselov}},\ }\href {\doibase
  10.1063/5.0048191} {\bibfield  {journal} {\bibinfo  {journal} {Applied
  Physics Letters}\ }\textbf {\bibinfo {volume} {118}},\ \bibinfo {pages}
  {183106} (\bibinfo {year} {2021})}\BibitemShut {NoStop}%
\bibitem [{\citenamefont {Greenaway}\ \emph {et~al.}(2015)\citenamefont
  {Greenaway}, \citenamefont {Vdovin}, \citenamefont {Mishchenko},
  \citenamefont {Makarovsky}, \citenamefont {Patan{\`e}}, \citenamefont
  {Wallbank}, \citenamefont {Cao}, \citenamefont {Kretinin}, \citenamefont
  {Zhu}, \citenamefont {Morozov}, \citenamefont {Fal'ko}, \citenamefont
  {Novoselov}, \citenamefont {Geim}, \citenamefont {Fromhold},\ and\
  \citenamefont {Eaves}}]{Greenaway2015}%
  \BibitemOpen
  \bibfield  {author} {\bibinfo {author} {\bibfnamefont {M.~T.}\ \bibnamefont
  {Greenaway}}, \bibinfo {author} {\bibfnamefont {E.~E.}\ \bibnamefont
  {Vdovin}}, \bibinfo {author} {\bibfnamefont {A.}~\bibnamefont {Mishchenko}},
  \bibinfo {author} {\bibfnamefont {O.}~\bibnamefont {Makarovsky}}, \bibinfo
  {author} {\bibfnamefont {A.}~\bibnamefont {Patan{\`e}}}, \bibinfo {author}
  {\bibfnamefont {J.~R.}\ \bibnamefont {Wallbank}}, \bibinfo {author}
  {\bibfnamefont {Y.}~\bibnamefont {Cao}}, \bibinfo {author} {\bibfnamefont
  {A.~V.}\ \bibnamefont {Kretinin}}, \bibinfo {author} {\bibfnamefont {M.~J.}\
  \bibnamefont {Zhu}}, \bibinfo {author} {\bibfnamefont {S.~V.}\ \bibnamefont
  {Morozov}}, \bibinfo {author} {\bibfnamefont {V.~I.}\ \bibnamefont {Fal'ko}},
  \bibinfo {author} {\bibfnamefont {K.~S.}\ \bibnamefont {Novoselov}}, \bibinfo
  {author} {\bibfnamefont {A.~K.}\ \bibnamefont {Geim}}, \bibinfo {author}
  {\bibfnamefont {T.~M.}\ \bibnamefont {Fromhold}}, \ and\ \bibinfo {author}
  {\bibfnamefont {L.}~\bibnamefont {Eaves}},\ }\href {\doibase
  10.1038/nphys3507} {\bibfield  {journal} {\bibinfo  {journal} {Nature
  Physics}\ }\textbf {\bibinfo {volume} {11}},\ \bibinfo {pages} {1057}
  (\bibinfo {year} {2015})}\BibitemShut {NoStop}%
\bibitem [{\citenamefont {Wallbank}\ \emph {et~al.}(2016)\citenamefont
  {Wallbank}, \citenamefont {Ghazaryan}, \citenamefont {Misra}, \citenamefont
  {Cao}, \citenamefont {Tu}, \citenamefont {Piot}, \citenamefont {Potemski},
  \citenamefont {Pezzini}, \citenamefont {Wiedmann}, \citenamefont {Zeitler},
  \citenamefont {Lane}, \citenamefont {Morozov}, \citenamefont {Greenaway},
  \citenamefont {Eaves}, \citenamefont {Geim}, \citenamefont {Fal'ko},
  \citenamefont {Novoselov},\ and\ \citenamefont {Mishchenko}}]{Wallbank2016}%
  \BibitemOpen
  \bibfield  {author} {\bibinfo {author} {\bibfnamefont {J.~R.}\ \bibnamefont
  {Wallbank}}, \bibinfo {author} {\bibfnamefont {D.}~\bibnamefont {Ghazaryan}},
  \bibinfo {author} {\bibfnamefont {A.}~\bibnamefont {Misra}}, \bibinfo
  {author} {\bibfnamefont {Y.}~\bibnamefont {Cao}}, \bibinfo {author}
  {\bibfnamefont {J.~S.}\ \bibnamefont {Tu}}, \bibinfo {author} {\bibfnamefont
  {B.~A.}\ \bibnamefont {Piot}}, \bibinfo {author} {\bibfnamefont
  {M.}~\bibnamefont {Potemski}}, \bibinfo {author} {\bibfnamefont
  {S.}~\bibnamefont {Pezzini}}, \bibinfo {author} {\bibfnamefont
  {S.}~\bibnamefont {Wiedmann}}, \bibinfo {author} {\bibfnamefont
  {U.}~\bibnamefont {Zeitler}}, \bibinfo {author} {\bibfnamefont {T.~L.~M.}\
  \bibnamefont {Lane}}, \bibinfo {author} {\bibfnamefont {S.~V.}\ \bibnamefont
  {Morozov}}, \bibinfo {author} {\bibfnamefont {M.~T.}\ \bibnamefont
  {Greenaway}}, \bibinfo {author} {\bibfnamefont {L.}~\bibnamefont {Eaves}},
  \bibinfo {author} {\bibfnamefont {A.~K.}\ \bibnamefont {Geim}}, \bibinfo
  {author} {\bibfnamefont {V.~I.}\ \bibnamefont {Fal'ko}}, \bibinfo {author}
  {\bibfnamefont {K.~S.}\ \bibnamefont {Novoselov}}, \ and\ \bibinfo {author}
  {\bibfnamefont {A.}~\bibnamefont {Mishchenko}},\ }\href {\doibase
  10.1126/science.aaf4621} {\bibfield  {journal} {\bibinfo  {journal}
  {Science}\ }\textbf {\bibinfo {volume} {353}},\ \bibinfo {pages} {575}
  (\bibinfo {year} {2016})}\BibitemShut {NoStop}%
\bibitem [{\citenamefont {McCann}(2006)}]{McCann2006}%
  \BibitemOpen
  \bibfield  {author} {\bibinfo {author} {\bibfnamefont {E.}~\bibnamefont
  {McCann}},\ }\href {\doibase 10.1103/PhysRevB.74.161403} {\bibfield
  {journal} {\bibinfo  {journal} {Phys. Rev. B}\ }\textbf {\bibinfo {volume}
  {74}},\ \bibinfo {pages} {161403} (\bibinfo {year} {2006})}\BibitemShut
  {NoStop}%
\bibitem [{\citenamefont {Ohta}\ \emph {et~al.}(2006)\citenamefont {Ohta},
  \citenamefont {Bostwick}, \citenamefont {Seyller}, \citenamefont {Horn},\
  and\ \citenamefont {Rotenberg}}]{Ohta2006}%
  \BibitemOpen
  \bibfield  {author} {\bibinfo {author} {\bibfnamefont {T.}~\bibnamefont
  {Ohta}}, \bibinfo {author} {\bibfnamefont {A.}~\bibnamefont {Bostwick}},
  \bibinfo {author} {\bibfnamefont {T.}~\bibnamefont {Seyller}}, \bibinfo
  {author} {\bibfnamefont {K.}~\bibnamefont {Horn}}, \ and\ \bibinfo {author}
  {\bibfnamefont {E.}~\bibnamefont {Rotenberg}},\ }\href {\doibase
  10.1126/science.1130681} {\bibfield  {journal} {\bibinfo  {journal}
  {Science}\ }\textbf {\bibinfo {volume} {313}},\ \bibinfo {pages} {951}
  (\bibinfo {year} {2006})}\BibitemShut {NoStop}%
\bibitem [{\citenamefont {Min}\ \emph {et~al.}(2007)\citenamefont {Min},
  \citenamefont {Sahu}, \citenamefont {Banerjee},\ and\ \citenamefont
  {MacDonald}}]{Min2007}%
  \BibitemOpen
  \bibfield  {author} {\bibinfo {author} {\bibfnamefont {H.}~\bibnamefont
  {Min}}, \bibinfo {author} {\bibfnamefont {B.}~\bibnamefont {Sahu}}, \bibinfo
  {author} {\bibfnamefont {S.~K.}\ \bibnamefont {Banerjee}}, \ and\ \bibinfo
  {author} {\bibfnamefont {A.~H.}\ \bibnamefont {MacDonald}},\ }\href {\doibase
  10.1103/PhysRevB.75.155115} {\bibfield  {journal} {\bibinfo  {journal} {Phys.
  Rev. B}\ }\textbf {\bibinfo {volume} {75}},\ \bibinfo {pages} {155115}
  (\bibinfo {year} {2007})}\BibitemShut {NoStop}%
\bibitem [{\citenamefont {Zhang}\ \emph {et~al.}(2009)\citenamefont {Zhang},
  \citenamefont {Tang}, \citenamefont {Girit}, \citenamefont {Hao},
  \citenamefont {Martin}, \citenamefont {Zettl}, \citenamefont {Crommie},
  \citenamefont {Shen},\ and\ \citenamefont {Wang}}]{Zhang2009}%
  \BibitemOpen
  \bibfield  {author} {\bibinfo {author} {\bibfnamefont {Y.}~\bibnamefont
  {Zhang}}, \bibinfo {author} {\bibfnamefont {T.-T.}\ \bibnamefont {Tang}},
  \bibinfo {author} {\bibfnamefont {C.}~\bibnamefont {Girit}}, \bibinfo
  {author} {\bibfnamefont {Z.}~\bibnamefont {Hao}}, \bibinfo {author}
  {\bibfnamefont {M.~C.}\ \bibnamefont {Martin}}, \bibinfo {author}
  {\bibfnamefont {A.}~\bibnamefont {Zettl}}, \bibinfo {author} {\bibfnamefont
  {M.~F.}\ \bibnamefont {Crommie}}, \bibinfo {author} {\bibfnamefont {Y.~R.}\
  \bibnamefont {Shen}}, \ and\ \bibinfo {author} {\bibfnamefont
  {F.}~\bibnamefont {Wang}},\ }\href {\doibase 10.1038/nature08105} {\bibfield
  {journal} {\bibinfo  {journal} {Nature}\ }\textbf {\bibinfo {volume} {459}},\
  \bibinfo {pages} {820} (\bibinfo {year} {2009})}\BibitemShut {NoStop}%
\bibitem [{\citenamefont {Yu}, \citenamefont {Stewart},\ and\ \citenamefont
  {Tiwari}(2008)}]{Yu2008}%
  \BibitemOpen
  \bibfield  {author} {\bibinfo {author} {\bibfnamefont {E.~K.}\ \bibnamefont
  {Yu}}, \bibinfo {author} {\bibfnamefont {D.~A.}\ \bibnamefont {Stewart}}, \
  and\ \bibinfo {author} {\bibfnamefont {S.}~\bibnamefont {Tiwari}},\ }\href
  {\doibase 10.1103/PhysRevB.77.195406} {\bibfield  {journal} {\bibinfo
  {journal} {Phys. Rev. B}\ }\textbf {\bibinfo {volume} {77}},\ \bibinfo
  {pages} {195406} (\bibinfo {year} {2008})}\BibitemShut {NoStop}%
\bibitem [{\citenamefont {Ramasubramaniam}, \citenamefont {Naveh},\ and\
  \citenamefont {Towe}(2011)}]{Ramasubramaniam2011}%
  \BibitemOpen
  \bibfield  {author} {\bibinfo {author} {\bibfnamefont {A.}~\bibnamefont
  {Ramasubramaniam}}, \bibinfo {author} {\bibfnamefont {D.}~\bibnamefont
  {Naveh}}, \ and\ \bibinfo {author} {\bibfnamefont {E.}~\bibnamefont {Towe}},\
  }\href {\doibase 10.1021/nl1039499} {\bibfield  {journal} {\bibinfo
  {journal} {Nano Letters}\ }\textbf {\bibinfo {volume} {11}},\ \bibinfo
  {pages} {1070} (\bibinfo {year} {2011})}\BibitemShut {NoStop}%
\bibitem [{\citenamefont {Kim}\ \emph {et~al.}(2013)\citenamefont {Kim},
  \citenamefont {Kim}, \citenamefont {Walter}, \citenamefont {Seyller},
  \citenamefont {Yeom}, \citenamefont {Rotenberg},\ and\ \citenamefont
  {Bostwick}}]{Kim2013}%
  \BibitemOpen
  \bibfield  {author} {\bibinfo {author} {\bibfnamefont {K.~S.}\ \bibnamefont
  {Kim}}, \bibinfo {author} {\bibfnamefont {T.-H.}\ \bibnamefont {Kim}},
  \bibinfo {author} {\bibfnamefont {A.~L.}\ \bibnamefont {Walter}}, \bibinfo
  {author} {\bibfnamefont {T.}~\bibnamefont {Seyller}}, \bibinfo {author}
  {\bibfnamefont {H.~W.}\ \bibnamefont {Yeom}}, \bibinfo {author}
  {\bibfnamefont {E.}~\bibnamefont {Rotenberg}}, \ and\ \bibinfo {author}
  {\bibfnamefont {A.}~\bibnamefont {Bostwick}},\ }\href {\doibase
  10.1103/PhysRevLett.110.036804} {\bibfield  {journal} {\bibinfo  {journal}
  {Phys. Rev. Lett.}\ }\textbf {\bibinfo {volume} {110}},\ \bibinfo {pages}
  {036804} (\bibinfo {year} {2013})}\BibitemShut {NoStop}%
\bibitem [{\citenamefont {Joucken}\ \emph {et~al.}(2021)\citenamefont
  {Joucken}, \citenamefont {Bena}, \citenamefont {Ge}, \citenamefont
  {Quezada-Lopez}, \citenamefont {Ducastelle}, \citenamefont {Tanagushi},
  \citenamefont {Watanabe},\ and\ \citenamefont {Velasco}}]{Joucken2021}%
  \BibitemOpen
  \bibfield  {author} {\bibinfo {author} {\bibfnamefont {F.}~\bibnamefont
  {Joucken}}, \bibinfo {author} {\bibfnamefont {C.}~\bibnamefont {Bena}},
  \bibinfo {author} {\bibfnamefont {Z.}~\bibnamefont {Ge}}, \bibinfo {author}
  {\bibfnamefont {E.~A.}\ \bibnamefont {Quezada-Lopez}}, \bibinfo {author}
  {\bibfnamefont {F.~m.~c.}\ \bibnamefont {Ducastelle}}, \bibinfo {author}
  {\bibfnamefont {T.}~\bibnamefont {Tanagushi}}, \bibinfo {author}
  {\bibfnamefont {K.}~\bibnamefont {Watanabe}}, \ and\ \bibinfo {author}
  {\bibfnamefont {J.}~\bibnamefont {Velasco}},\ }\href {\doibase
  10.1103/PhysRevLett.127.106401} {\bibfield  {journal} {\bibinfo  {journal}
  {Phys. Rev. Lett.}\ }\textbf {\bibinfo {volume} {127}},\ \bibinfo {pages}
  {106401} (\bibinfo {year} {2021})}\BibitemShut {NoStop}%
\bibitem [{\citenamefont {Li}\ \emph {et~al.}(2017)\citenamefont {Li},
  \citenamefont {Taniguchi}, \citenamefont {Watanabe}, \citenamefont {Hone},\
  and\ \citenamefont {Dean}}]{Li2017}%
  \BibitemOpen
  \bibfield  {author} {\bibinfo {author} {\bibfnamefont {J.~I.~A.}\
  \bibnamefont {Li}}, \bibinfo {author} {\bibfnamefont {T.}~\bibnamefont
  {Taniguchi}}, \bibinfo {author} {\bibfnamefont {K.}~\bibnamefont {Watanabe}},
  \bibinfo {author} {\bibfnamefont {J.}~\bibnamefont {Hone}}, \ and\ \bibinfo
  {author} {\bibfnamefont {C.~R.}\ \bibnamefont {Dean}},\ }\href {\doibase
  10.1038/nphys4140} {\bibfield  {journal} {\bibinfo  {journal} {Nature
  Physics}\ }\textbf {\bibinfo {volume} {13}},\ \bibinfo {pages} {751}
  (\bibinfo {year} {2017})}\BibitemShut {NoStop}%
\bibitem [{\citenamefont {Zeng}\ \emph {et~al.}(2025)\citenamefont {Zeng},
  \citenamefont {Sun}, \citenamefont {Zhang}, \citenamefont {Nguyen},
  \citenamefont {Shi}, \citenamefont {Okounkova}, \citenamefont {Watanabe},
  \citenamefont {Taniguchi}, \citenamefont {Hone}, \citenamefont {Dean},\ and\
  \citenamefont {Li}}]{Zeng2025}%
  \BibitemOpen
  \bibfield  {author} {\bibinfo {author} {\bibfnamefont {Y.}~\bibnamefont
  {Zeng}}, \bibinfo {author} {\bibfnamefont {D.}~\bibnamefont {Sun}}, \bibinfo
  {author} {\bibfnamefont {N.~J.}\ \bibnamefont {Zhang}}, \bibinfo {author}
  {\bibfnamefont {R.~Q.}\ \bibnamefont {Nguyen}}, \bibinfo {author}
  {\bibfnamefont {Q.}~\bibnamefont {Shi}}, \bibinfo {author} {\bibfnamefont
  {A.}~\bibnamefont {Okounkova}}, \bibinfo {author} {\bibfnamefont
  {K.}~\bibnamefont {Watanabe}}, \bibinfo {author} {\bibfnamefont
  {T.}~\bibnamefont {Taniguchi}}, \bibinfo {author} {\bibfnamefont
  {J.}~\bibnamefont {Hone}}, \bibinfo {author} {\bibfnamefont {C.~R.}\
  \bibnamefont {Dean}}, \ and\ \bibinfo {author} {\bibfnamefont {J.~I.~A.}\
  \bibnamefont {Li}},\ }\href {https://arxiv.org/abs/2306.16995} {\enquote
  {\bibinfo {title} {Evidence for a superfluid-to-solid transition of bilayer
  excitons},}\ } (\bibinfo {year} {2025}),\ \Eprint
  {http://arxiv.org/abs/2306.16995} {arXiv:2306.16995 [cond-mat.mes-hall]}
  \BibitemShut {NoStop}%
\bibitem [{\citenamefont {de~la Barrera}\ \emph {et~al.}(2022)\citenamefont
  {de~la Barrera}, \citenamefont {Aronson}, \citenamefont {Zheng},
  \citenamefont {Watanabe}, \citenamefont {Taniguchi}, \citenamefont {Ma},
  \citenamefont {Jarillo-Herrero},\ and\ \citenamefont
  {Ashoori}}]{delaBarrera2022}%
  \BibitemOpen
  \bibfield  {author} {\bibinfo {author} {\bibfnamefont {S.~C.}\ \bibnamefont
  {de~la Barrera}}, \bibinfo {author} {\bibfnamefont {S.}~\bibnamefont
  {Aronson}}, \bibinfo {author} {\bibfnamefont {Z.}~\bibnamefont {Zheng}},
  \bibinfo {author} {\bibfnamefont {K.}~\bibnamefont {Watanabe}}, \bibinfo
  {author} {\bibfnamefont {T.}~\bibnamefont {Taniguchi}}, \bibinfo {author}
  {\bibfnamefont {Q.}~\bibnamefont {Ma}}, \bibinfo {author} {\bibfnamefont
  {P.}~\bibnamefont {Jarillo-Herrero}}, \ and\ \bibinfo {author} {\bibfnamefont
  {R.}~\bibnamefont {Ashoori}},\ }\href {\doibase 10.1038/s41567-022-01616-w}
  {\bibfield  {journal} {\bibinfo  {journal} {Nature Physics}\ }\textbf
  {\bibinfo {volume} {18}},\ \bibinfo {pages} {771} (\bibinfo {year}
  {2022})}\BibitemShut {NoStop}%
\bibitem [{\citenamefont {Seiler}\ \emph {et~al.}(2022)\citenamefont {Seiler},
  \citenamefont {Geisenhof}, \citenamefont {Winterer}, \citenamefont
  {Watanabe}, \citenamefont {Taniguchi}, \citenamefont {Xu}, \citenamefont
  {Zhang},\ and\ \citenamefont {Weitz}}]{Seiler2022}%
  \BibitemOpen
  \bibfield  {author} {\bibinfo {author} {\bibfnamefont {A.~M.}\ \bibnamefont
  {Seiler}}, \bibinfo {author} {\bibfnamefont {F.~R.}\ \bibnamefont
  {Geisenhof}}, \bibinfo {author} {\bibfnamefont {F.}~\bibnamefont {Winterer}},
  \bibinfo {author} {\bibfnamefont {K.}~\bibnamefont {Watanabe}}, \bibinfo
  {author} {\bibfnamefont {T.}~\bibnamefont {Taniguchi}}, \bibinfo {author}
  {\bibfnamefont {T.}~\bibnamefont {Xu}}, \bibinfo {author} {\bibfnamefont
  {F.}~\bibnamefont {Zhang}}, \ and\ \bibinfo {author} {\bibfnamefont {R.~T.}\
  \bibnamefont {Weitz}},\ }\href {\doibase 10.1038/s41586-022-04937-1}
  {\bibfield  {journal} {\bibinfo  {journal} {Nature}\ }\textbf {\bibinfo
  {volume} {608}},\ \bibinfo {pages} {298} (\bibinfo {year}
  {2022})}\BibitemShut {NoStop}%
\bibitem [{\citenamefont {Vdovin}\ \emph {et~al.}(2024)\citenamefont {Vdovin},
  \citenamefont {Khanin}, \citenamefont {Morozov}, \citenamefont {Kashchenko},
  \citenamefont {Sokolik},\ and\ \citenamefont {Novoselov}}]{Vdovin2024}%
  \BibitemOpen
  \bibfield  {author} {\bibinfo {author} {\bibfnamefont {E.~E.}\ \bibnamefont
  {Vdovin}}, \bibinfo {author} {\bibfnamefont {Y.~N.}\ \bibnamefont {Khanin}},
  \bibinfo {author} {\bibfnamefont {S.~V.}\ \bibnamefont {Morozov}}, \bibinfo
  {author} {\bibfnamefont {M.~A.}\ \bibnamefont {Kashchenko}}, \bibinfo
  {author} {\bibfnamefont {A.~A.}\ \bibnamefont {Sokolik}}, \ and\ \bibinfo
  {author} {\bibfnamefont {K.~S.}\ \bibnamefont {Novoselov}},\ }\href {\doibase
  10.1134/S0021364024604019} {\bibfield  {journal} {\bibinfo  {journal} {JETP
  Letters}\ }\textbf {\bibinfo {volume} {120}},\ \bibinfo {pages} {854}
  (\bibinfo {year} {2024})}\BibitemShut {NoStop}%
\bibitem [{\citenamefont {Mucha-Kruczyński}, \citenamefont {McCann},\ and\
  \citenamefont {Fal'ko}(2010)}]{Mucha-Kruczynski2010}%
  \BibitemOpen
  \bibfield  {author} {\bibinfo {author} {\bibfnamefont {M.}~\bibnamefont
  {Mucha-Kruczyński}}, \bibinfo {author} {\bibfnamefont {E.}~\bibnamefont
  {McCann}}, \ and\ \bibinfo {author} {\bibfnamefont {V.~I.}\ \bibnamefont
  {Fal'ko}},\ }\href {\doibase 10.1088/0268-1242/25/3/033001} {\ \textbf
  {\bibinfo {volume} {25}},\ \bibinfo {pages} {033001} (\bibinfo {year}
  {2010})}\BibitemShut {NoStop}%
\bibitem [{\citenamefont {Zou}, \citenamefont {Hong},\ and\ \citenamefont
  {Zhu}(2011)}]{Zou2011}%
  \BibitemOpen
  \bibfield  {author} {\bibinfo {author} {\bibfnamefont {K.}~\bibnamefont
  {Zou}}, \bibinfo {author} {\bibfnamefont {X.}~\bibnamefont {Hong}}, \ and\
  \bibinfo {author} {\bibfnamefont {J.}~\bibnamefont {Zhu}},\ }\href {\doibase
  10.1103/PhysRevB.84.085408} {\bibfield  {journal} {\bibinfo  {journal} {Phys.
  Rev. B}\ }\textbf {\bibinfo {volume} {84}},\ \bibinfo {pages} {085408}
  (\bibinfo {year} {2011})}\BibitemShut {NoStop}%
\bibitem [{\citenamefont {McCann}\ and\ \citenamefont
  {Koshino}(2013)}]{McCann2013}%
  \BibitemOpen
  \bibfield  {author} {\bibinfo {author} {\bibfnamefont {E.}~\bibnamefont
  {McCann}}\ and\ \bibinfo {author} {\bibfnamefont {M.}~\bibnamefont
  {Koshino}},\ }\href {\doibase 10.1088/0034-4885/76/5/056503} {\bibfield
  {journal} {\bibinfo  {journal} {Reports on Progress in Physics}\ }\textbf
  {\bibinfo {volume} {76}},\ \bibinfo {pages} {056503} (\bibinfo {year}
  {2013})}\BibitemShut {NoStop}%
\bibitem [{\citenamefont {Lane}, \citenamefont {Wallbank},\ and\ \citenamefont
  {Fal'ko}(2015)}]{Lane2015}%
  \BibitemOpen
  \bibfield  {author} {\bibinfo {author} {\bibfnamefont {T.~L.~M.}\
  \bibnamefont {Lane}}, \bibinfo {author} {\bibfnamefont {J.~R.}\ \bibnamefont
  {Wallbank}}, \ and\ \bibinfo {author} {\bibfnamefont {V.~I.}\ \bibnamefont
  {Fal'ko}},\ }\href {\doibase 10.1063/1.4935988} {\bibfield  {journal}
  {\bibinfo  {journal} {Applied Physics Letters}\ }\textbf {\bibinfo {volume}
  {107}},\ \bibinfo {pages} {203506} (\bibinfo {year} {2015})}\BibitemShut
  {NoStop}%
\bibitem [{\citenamefont {Thompson}, \citenamefont {Leech},\ and\ \citenamefont
  {Mucha-Kruczy\ifmmode~\acute{n}\else \'{n}\fi{}ski}(2019)}]{Thompson2019}%
  \BibitemOpen
  \bibfield  {author} {\bibinfo {author} {\bibfnamefont {J.~J.~P.}\
  \bibnamefont {Thompson}}, \bibinfo {author} {\bibfnamefont {D.~J.}\
  \bibnamefont {Leech}}, \ and\ \bibinfo {author} {\bibfnamefont
  {M.}~\bibnamefont {Mucha-Kruczy\ifmmode~\acute{n}\else \'{n}\fi{}ski}},\
  }\href {\doibase 10.1103/PhysRevB.99.085420} {\bibfield  {journal} {\bibinfo
  {journal} {Phys. Rev. B}\ }\textbf {\bibinfo {volume} {99}},\ \bibinfo
  {pages} {085420} (\bibinfo {year} {2019})}\BibitemShut {NoStop}%
\bibitem [{\citenamefont {Yankowitz}\ \emph {et~al.}(2019)\citenamefont
  {Yankowitz}, \citenamefont {Ma}, \citenamefont {Jarillo-Herrero},\ and\
  \citenamefont {LeRoy}}]{Yankowitz2019}%
  \BibitemOpen
  \bibfield  {author} {\bibinfo {author} {\bibfnamefont {M.}~\bibnamefont
  {Yankowitz}}, \bibinfo {author} {\bibfnamefont {Q.}~\bibnamefont {Ma}},
  \bibinfo {author} {\bibfnamefont {P.}~\bibnamefont {Jarillo-Herrero}}, \ and\
  \bibinfo {author} {\bibfnamefont {B.~J.}\ \bibnamefont {LeRoy}},\ }\href
  {\doibase 10.1038/s42254-018-0016-0} {\bibfield  {journal} {\bibinfo
  {journal} {Nature Reviews Physics}\ }\textbf {\bibinfo {volume} {1}},\
  \bibinfo {pages} {112} (\bibinfo {year} {2019})}\BibitemShut {NoStop}%
\bibitem [{\citenamefont {Inbar}\ \emph {et~al.}(2023)\citenamefont {Inbar},
  \citenamefont {Birkbeck}, \citenamefont {Xiao}, \citenamefont {Taniguchi},
  \citenamefont {Watanabe}, \citenamefont {Yan}, \citenamefont {Oreg},
  \citenamefont {Stern}, \citenamefont {Berg},\ and\ \citenamefont
  {Ilani}}]{Inbar2023}%
  \BibitemOpen
  \bibfield  {author} {\bibinfo {author} {\bibfnamefont {A.}~\bibnamefont
  {Inbar}}, \bibinfo {author} {\bibfnamefont {J.}~\bibnamefont {Birkbeck}},
  \bibinfo {author} {\bibfnamefont {J.}~\bibnamefont {Xiao}}, \bibinfo {author}
  {\bibfnamefont {T.}~\bibnamefont {Taniguchi}}, \bibinfo {author}
  {\bibfnamefont {K.}~\bibnamefont {Watanabe}}, \bibinfo {author}
  {\bibfnamefont {B.}~\bibnamefont {Yan}}, \bibinfo {author} {\bibfnamefont
  {Y.}~\bibnamefont {Oreg}}, \bibinfo {author} {\bibfnamefont {A.}~\bibnamefont
  {Stern}}, \bibinfo {author} {\bibfnamefont {E.}~\bibnamefont {Berg}}, \ and\
  \bibinfo {author} {\bibfnamefont {S.}~\bibnamefont {Ilani}},\ }\href
  {\doibase 10.1038/s41586-022-05685-y} {\bibfield  {journal} {\bibinfo
  {journal} {Nature}\ }\textbf {\bibinfo {volume} {614}},\ \bibinfo {pages}
  {682} (\bibinfo {year} {2023})}\BibitemShut {NoStop}%
\end{thebibliography}%

\end{document}


\renewcommand{\theequation}{S\arabic{equation}}
\renewcommand{\thefigure}{S\arabic{figure}}

\title{Supplementary Material for ``Probing the features of electron dispersion by tunneling between slightly twisted bilayer graphene sheets''}

\author{Alexey A. Sokolik}
\email{asokolik@hse.ru}
\affiliation{Institute for Spectroscopy, Russian Academy of Sciences, 108840 Troitsk, Moscow, Russia}
\affiliation{National Research University Higher School of Economics, 109028 Moscow, Russia}

\author{Azat F. Aminov}
\affiliation{National Research University Higher School of Economics, 109028 Moscow, Russia}
\affiliation{Institute of Microelectronics Technology and High Purity Materials, Russian Academy of Sciences, 142432 Chernogolovka, Russia}

\author{Evgenii E. Vdovin}
\author{Yurii N. Khanin}
\affiliation{Institute of Microelectronics Technology and High Purity Materials, Russian Academy of Sciences, 142432 Chernogolovka, Russia}

\author{Mikhail A. Kashchenko}
\affiliation{Programmable Functional Materials Lab, Center for Neurophysics and Neuromorphic
Technologies, 127495 Moscow, Russia}
\affiliation{Moscow Center for Advanced Studies, 123592 Moscow, Russia}

\author{Denis A. Bandurin}
\affiliation{Department of Materials Science and Engineering, National University of Singapore, 117575 Singapore, Singapore}
\affiliation{Programmable Functional Materials Lab, Center for Neurophysics and Neuromorphic
Technologies, 127495 Moscow, Russia}
\affiliation{Institute for Functional Intelligent Materials, National University of Singapore, 117544 Singapore, Singapore}

\author{Davit A. Ghazaryan}
\affiliation{Laboratory of Advanced Functional Materials, Yerevan State University, 0025 Yerevan, Armenia}
\affiliation{Institute for Functional Intelligent Materials, National University of Singapore, 117544 Singapore, Singapore}

\author{Sergey V. Morozov}
\affiliation{Institute of Microelectronics Technology and High Purity Materials, Russian Academy of Sciences, 142432 Chernogolovka, Russia}

\author{Kostya S. Novoselov}
\email{kostya@nus.edu.sg}
\affiliation{Institute for Functional Intelligent Materials, National University of Singapore, 117544 Singapore, Singapore}

\maketitle

\section{Sample preparation}

Two devices, Device 1 and Device 2, were fabricated using identical methods for the active layers. The process began with the mechanical exfoliation of Bernal bilayer graphene and hBN flakes onto a $\mbox{Si/SiO}_2$ substrate. The desired van der Waals heterostructure was assembled layer by layer (from top to bottom layer) using a PC/PDMS stamp, and finally transferred onto the target substrate. To ensure precise alignment with a near-zero twist angle, adjacent graphene flakes from the same exfoliation were used. The final stack for both samples consisted of a bottom graphene electrode, an hBN tunnel barrier, and a top graphene electrode. Bernal stacking of our bilayer graphene electrodes has been verified by measuring Raman spectra. The bottom gate differs in the two heterostructures: Device 1 used the silicon substrate separated by a 300 nm thick $\mbox{SiO}_2$ dielectric layer, while Device 2 used a graphite flake. In the final heterostructures, Device 1 had a twist angle of approximately $0.7^\circ$, while Device 2 had a twist angle of approximately $0.1^\circ$. In Device 2, the bottom BLG is twisted by about $1^\circ$ with respect to the lowermost hBN layer, and the top BLG is twisted by about $3^\circ$ with respect to the uppermost hBN layer.

\begin{figure}
\begin{center}
\includegraphics[width=0.77\columnwidth]{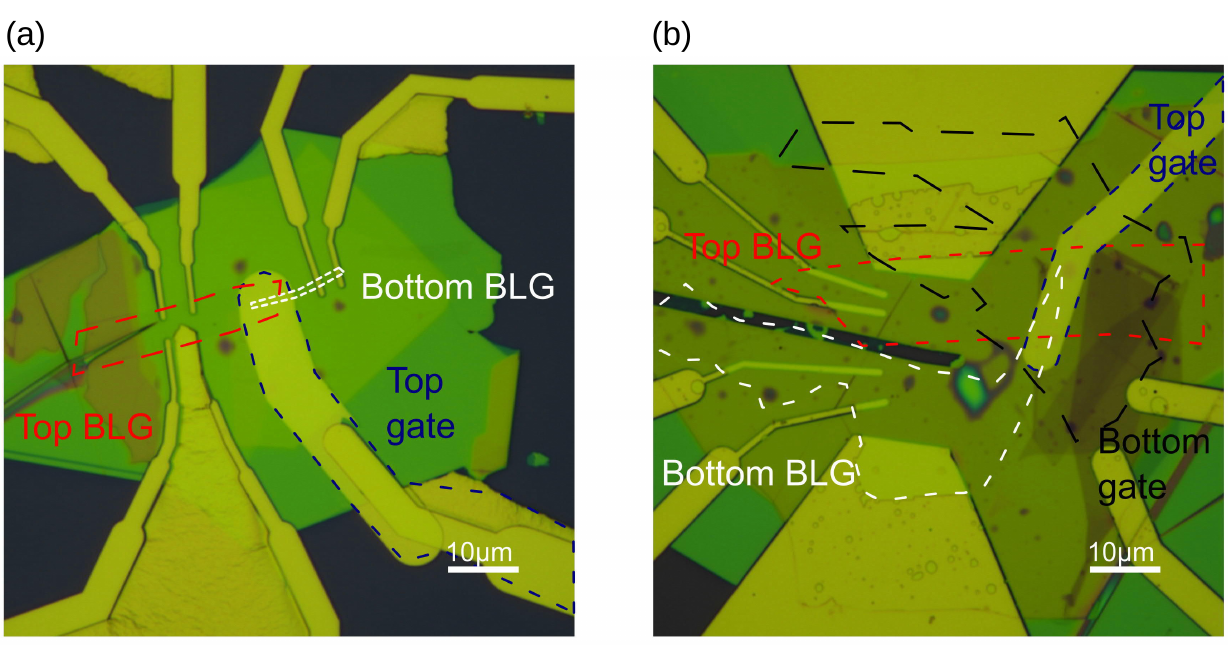}
\end{center}
\caption{\label{SI_Fig1}Photographs of (a) Device 1 and (b) Device 2.}
\end{figure}

The photographs of the devices are shown in Fig.~\ref{SI_Fig1}. Contacts to the graphene layers and the top gate were defined by electron beam lithography, followed by metal deposition Cr/Au and lift-off.  An $\mathrm{SF}_6$ RIE process was used to selectively etch the top hBN before metal deposition. The contact geometry allowed for four-probe measurements. Active areas of the devices (graphene flake overlap) were about $1.4\,\mu\mbox{m}^2$ for Device 1 and $18\,\mu\mbox{m}^2$ for Device 2.

\section{Electrostatic model}\label{sec_electrostatic}
	
\begin{figure}[b]
\begin{center}
\includegraphics[width=0.55\columnwidth]{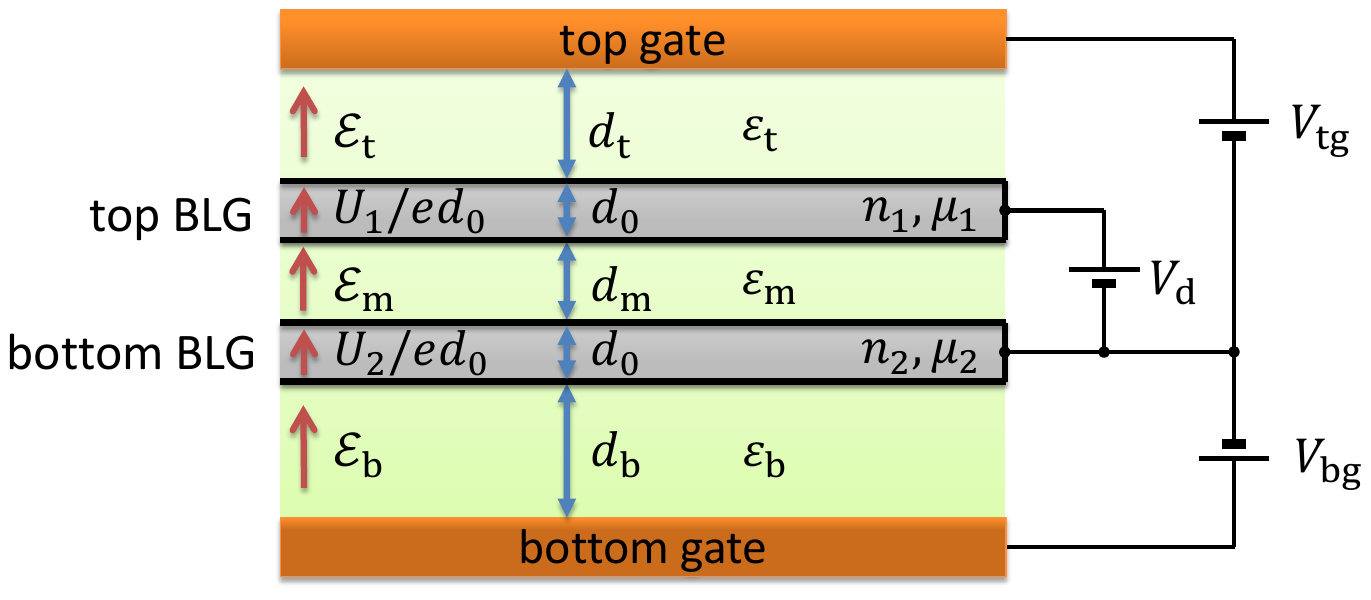}
\end{center}
\caption{\label{SI_Fig2}Electrostatic layout of the system, where electric fields $\mathcal{E}_i$, dielectric thicknesses $d_i$, and permittivities $\varepsilon_i$ are shown for each insulating layer. Top and bottom BLGs are characterized by the total electron densities $n_{1,2}$ and by the electron chemical potentials $\mu_{1,2}$.}
\end{figure}

To describe the electrostatics of the system, we use the Gauss law relating electron densities on top and bottom BLGs to electric fields in the insulating layers (see Fig.~\ref{SI_Fig2}):
\begin{equation}
4\pi en_1=-\varepsilon_\mathrm{t}\mathcal{E}_\mathrm{t}+\varepsilon_\mathrm{m}\mathcal{E}_\mathrm{m},\qquad4\pi en_2=-\varepsilon_\mathrm{m}\mathcal{E}_\mathrm{m}+\varepsilon_\mathrm{b}\mathcal{E}_\mathrm{b}.\label{SI_el1}
\end{equation}
The differences of Gauss laws for top and bottom graphene sublayers of each BLG provide connection with the internal fields $U_i/ed_0$:
\begin{equation}
4\pi e\Delta n_1=2U_1/ed_0-\varepsilon_\mathrm{t}\mathcal{E}_\mathrm{t}-\varepsilon_\mathrm{m}\mathcal{E}_\mathrm{m},\qquad4\pi e\Delta n_2=2U_2/ed_0-\varepsilon_\mathrm{b}\mathcal{E}_\mathrm{b}-\varepsilon_\mathrm{m}\mathcal{E}_\mathrm{m},\label{SI_el2}
\end{equation}
where $\Delta n_i$ is the difference of electron densities in top and bottom graphene sublayers of the $i$-th BLG. On the other hand, the difference of electrochemical potentials in each pair of contacts is equal to the voltage applied to it in equilibrium. For top BLG and top gate, bottom BLG and top BLG, bottom BLG and bottom gate we obtain, respectively,
\begin{equation}
\mu_1-ed_\mathrm{t}\mathcal{E}_\mathrm{t}-\frac{U_1}2=e(V_\mathrm{tg}-V_\mathrm{d}),\qquad\mu_2-\mu_1-ed_\mathrm{m}\mathcal{E}_\mathrm{m}-\frac{U_1+U_2}2=eV_\mathrm{d},\qquad\mu_2+ed_\mathrm{b}\mathcal{E}_\mathrm{b}+\frac{U_2}2=eV_\mathrm{bg}.\label{SI_el3}
\end{equation}
Here we have connected the differences of electrostatic potentials with electric fields and taken into account the potential differences $U_i$, which are symmetrically biasing the potentials of graphene sublayers of each BLG. Excluding the electric fields $\mathcal{E}_i$ from Eqs.~(\ref{SI_el1})--(\ref{SI_el3}), we obtain
\begin{equation}
\left\{\begin{array}{l}\displaystyle en_1=C_\mathrm{t}\left(V_\mathrm{tg}-V_\mathrm{d}-\frac{\mu_1}e+\frac{U_1}{2e}\right)-C_\mathrm{m}\left(V_\mathrm{d}+\frac{\mu_1}e-\frac{\mu_2}e+\frac{U_1}{2e}+\frac{U_2}{2e}\right),\\\\
\displaystyle en_2=C_\mathrm{b}\left(V_\mathrm{bg}-\frac{\mu_2}e-\frac{U_2}{2e}\right)+C_\mathrm{m}\left(V_\mathrm{d}+\frac{\mu_1}e-\frac{\mu_2}e+\frac{U_1}{2e}+\frac{U_2}{2e}\right),\\\\
\displaystyle 2C_0\frac{U_1}e=e\Delta n_1-C_\mathrm{t}\left(V_\mathrm{tg}-V_\mathrm{d}-\frac{\mu_1}e+\frac{U_1}{2e}\right)-C_\mathrm{m}\left(V_\mathrm{d}+\frac{\mu_1}e-\frac{\mu_2}e+\frac{U_1}{2e}+\frac{U_2}{2e}\right),\\\\
\displaystyle 2C_0\frac{U_2}e=e\Delta n_2+C_\mathrm{b}\left(V_\mathrm{bg}-\frac{\mu_2}e-\frac{U_2}{2e}\right)-C_\mathrm{m}\left(V_\mathrm{d}+\frac{\mu_1}e-\frac{\mu_2}e+\frac{U_1}{2e}+\frac{U_2}{2e}\right).\end{array}\right.\label{SI_el4}
\end{equation}
Here $C_j=\varepsilon_j/4\pi d_j$ are the geometric capacitances per unit area for three dielectric layers $j=\mathrm{t,m,b}$, and $C_0=1/4\pi d_0$ is the inter-sublayer capacitance per unit area of BLG. The second pair of equations (\ref{SI_el4}) imply that the external vertical electric field is self-consistently screened inside each BLG, because $U_i$ and $\Delta n_i$ are of the same sign. To close this system of equations, we express the chemical potentials $\mu_i$ and sublayer density imbalances $\Delta n_i$ through electron densities $n_i$ and induced gaps $U_i$. The chemical potentials are equal to electron energies at Fermi momenta,
\begin{equation}
\mu_i=E_{k_\mathrm{F}i,\mathrm{sgn}(n_i)},\qquad k_{\mathrm{F}i}=\sqrt{\pi|n_i|},
\end{equation}
were the electron dispersions $E_{k,\pm}$ in the conduction and valence bands are obtained by diagonalizing the Hamiltonian (see Eq.~(5) from the main text). The density imbalances can be calculated from the simple parabolic dispersion model \cite{McCann2013}:
\begin{equation}
\Delta n_i=\frac{U_in_\perp}{2\gamma_1}\ln\left\{\frac{|n_i|}{2n_\perp}+\frac12\sqrt{\left(\frac{n_i}{n_\perp}\right)^2+\left(\frac{U_i}{2\gamma_1}\right)^2}\right\},\label{SI_el6}
\end{equation}
where $n_\perp=\gamma_1^2/\pi\hbar^2v^2$. Assuming $V_\mathrm{d}=0$ and solving the system of equations (\ref{SI_el4})--(\ref{SI_el6}) for each pair $V_\mathrm{tg}$, $V_\mathrm{bg}$, we find the parameters $\mu_i$, $U_i$, which are used to calculate the tunneling conductance according to the formula (3) from the main text.

\section{Calculation parameters}

In our calculations, we use the parameters of electron dispersion, which are generally accepted for BLG \cite{McCann2013}:
\begin{equation}
\gamma_1=0.381\,\mbox{eV},\qquad v=10^6\,\mbox{m/s},\qquad v_4=0.05v.
\end{equation}
The asymmetry parameter $v_4=0.05v$ corresponds to the value $\gamma_4=0.05\gamma_0\approx0.15\,\mbox{eV}$ of the diagonal hopping integral in BLG breaking the electron-holy symmetry (where $\gamma_0\approx3\,\mbox{eV}$), which agrees with the range 0.04-0.15 eV of this parameter reported throughout the literature \cite{McCann2013}. However our fitting of the parameters does not allow for more accurate quantitative estimate of this asymmetry.

Temperature in the Fermi-Dirac distribution equals to 4.5 K, close to the experimental conditions, and the energy width of electron states is $\Gamma=30\,\mbox{K}$ for Device 1 and $\Gamma=10\,\mbox{K}$ for Device 2. These broadening parameters were chosen to reproduce the experimentally observable degrees of smoothness of the tunneling resonances and allowed tunneling regions. Being the phenomenological parameter, $\Gamma$ takes into account not only finite lifetime of the electron states, but also sample inhomogeneities and partial electron momentum nonconservation during the tunneling. Oscillations of the bias voltage $V_\mathrm{d}$ during the tunneling conductance measurements can also provide an additional smoothing of the energies in tunneling processes, which is additionally absorbed in $\Gamma$.

Parameters of the electrostatic model are fitted in order to reproduce as closely as possible the experimental voltage-density relationships:
\begin{align}
&\mbox{Device 1}:\quad d_\mathrm{t}=37.75\,\mbox{nm},\quad d_\mathrm{m}=2\,\mbox{nm},\quad
d_\mathrm{b}= 234.8\,\mbox{nm},\quad V_\mathrm{tg}^{(0)}=-1\,\mbox{V},\quad V_\mathrm{bg}^{(0)}=-5.8\,\mbox{V}.\\
&\mbox{Device 2}:\quad d_\mathrm{t}=40.4\,\mbox{nm},\quad d_\mathrm{m}=1.5\,\mbox{nm},\quad
d_\mathrm{b}=47.8\,\mbox{nm},\quad V_\mathrm{tg}^{(0)}=0.425\,\mbox{V},\quad V_\mathrm{bg}^{(0)}=-0.095\,\mbox{V}.
\end{align}
The distance between graphene sublayers of BLG is $d_0=3.35\,\mbox{\AA}$. The constant voltages $V_\mathrm{tg}^{(0)}$, $V_\mathrm{bg}^{(0)}$ are added to those in Eqs.~(\ref{SI_el4}) in order to compensate the experimental displacement of BLG charge neutrality points. All dielectric constants are equated to that of hexagonal boron nitride: $\varepsilon_\mathrm{t}=\varepsilon_\mathrm{m}=\varepsilon_\mathrm{b}=3.2$. Although this constant can vary depending on the thickness \cite{Laturia2018}, in our fitting procedure it is equivalent to changes of the fitting parameters $d_j$. Note that in Device 1 sample there is an additional $\mathrm{SiO}_2$ layer above the bottom gate, whose thickness and dielectric constant are absorbed in the effective thickness $d_\mathrm{b}= 234.8\,\mbox{nm}$ in our calculations.

\section{Normalized comparison of the tunneling conductance maps}

\begin{figure}[!t]
\begin{center}
\includegraphics[width=0.61\columnwidth]{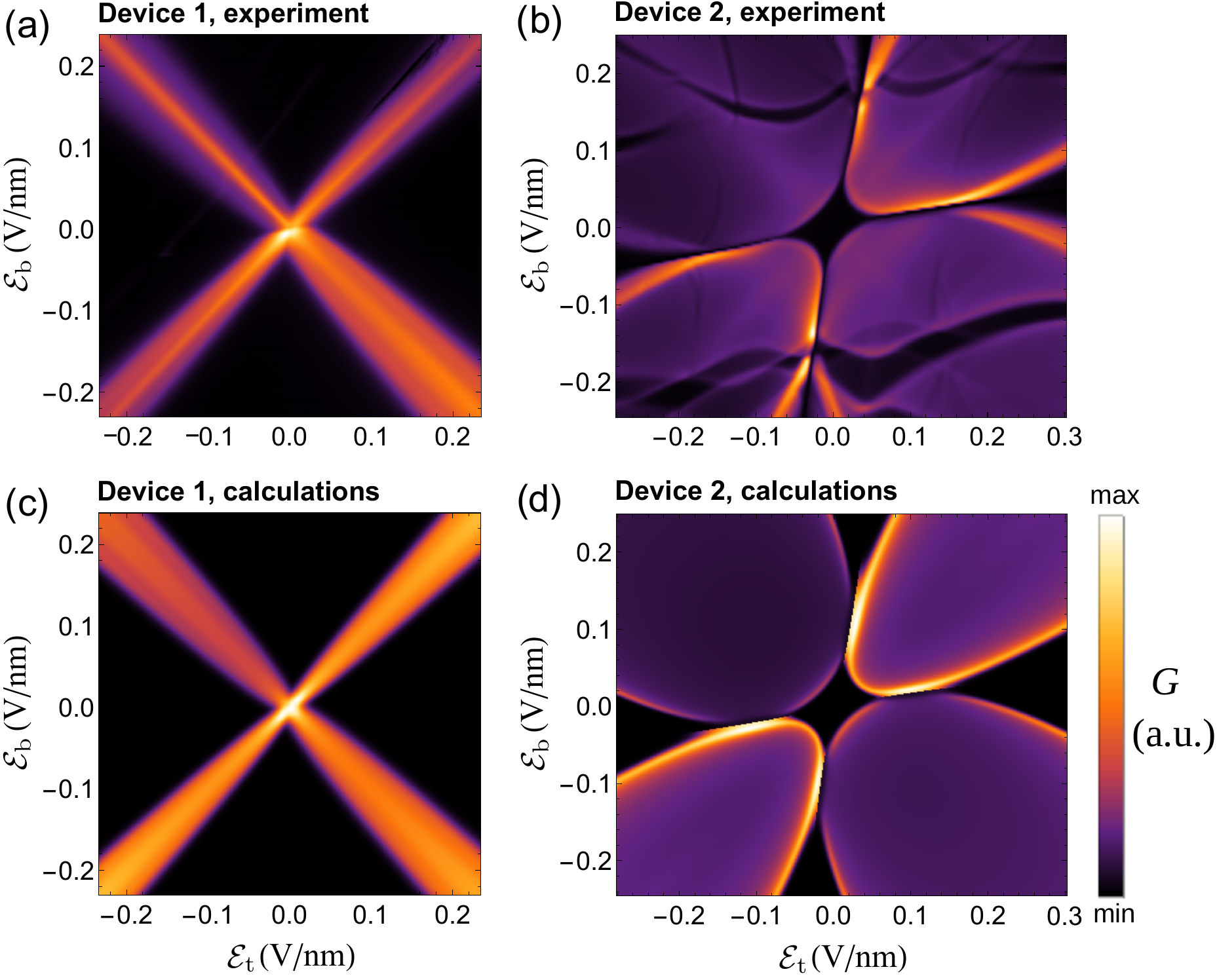}
\end{center}
\caption{\label{SI_Fig3}Maps of tunneling conductance (in arbitrary units) as functions of electric fields (\ref{el_fields}) in top and bottom dielectric layers for (a,c) Device 1 and (b,d) Device 2. The upper panels (a,b) show experimental results and the lower panels (c,d) show theoretical calculations.}
\end{figure}

In this section, we provide the tunneling conductance maps for Devices 1 and 2 in terms of electric fields instead of the gate voltages. The electric fields in top ($\mathcal{E}_\mathrm{t}$) and bottom ($\mathcal{E}_\mathrm{b}$) dielectric layers can be found from Eq.~(\ref{SI_el3}):
\begin{equation}
\mathcal{E}_\mathrm{t}=\frac{-e(V_\mathrm{tg}-V_\mathrm{d})+\mu_1-\frac12U_1}{ed_\mathrm{t}},\qquad\mathcal{E}_\mathrm{b}=\frac{eV_\mathrm{bg}-\mu_2-\frac12U_2}{ed_\mathrm{b}}.\label{el_fields}
\end{equation}
Using the electrostatic model described in Section~\ref{sec_electrostatic}, we recalculate experimental and theoretical maps and show them in Fig.~\ref{SI_Fig3}. Such picture provides normalized comparison of the maps for Device 1 and Device 2 by eliminating the differences in $d_\mathrm{t}$ and $d_\mathrm{b}$, because equal values of the electric fields correspond to approximate equal doping levels (see the first two equations of the system (\ref{SI_el4})), if we neglect a contribution of the weaker field $\mathcal{E}_\mathrm{m}$ between the BLGs.

\begin{figure}[!b]
\begin{center}
\includegraphics[width=0.85\columnwidth]{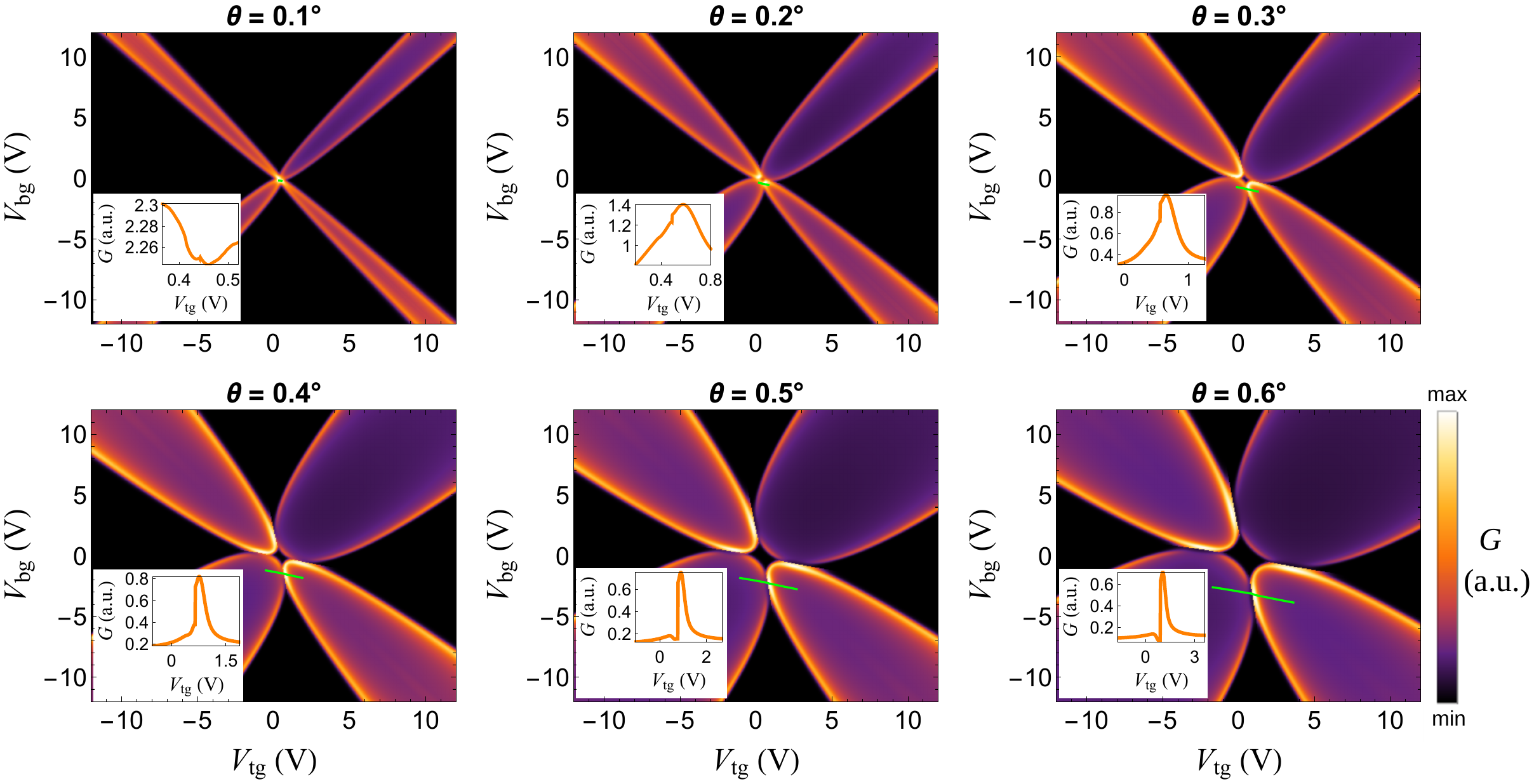}
\end{center}
\caption{\label{SI_Fig4}Maps of tunneling conductance (in arbitrary units) as functions of top and bottom gate voltages for Device 2, where the twist angle $\theta$ was changed to other values indicated above each diagram. Green curves show the cross-sections of the maps, and insets present calculated tunneling conductance along these cross-sections.}
\end{figure}

\section{Calculations for intermediate twist angles}

In this section, we present a series of calculated tunneling conductance maps for progressively increasing twist angles $\theta$ from $0.1^\circ$ to $0.6^\circ$, see Fig.~\ref{SI_Fig4}. All calculations were carried out with the parameters of Device 2, and only $\theta$ was changed with all other parameters remaining the same. One can notice the progressive widening of the allowed tunneling regions and appearance of the forbidden region in the center as $\theta$ increases.

We also take the example cross-sections of the tunneling conductance maps, where $n_2=\mathrm{const}$, and $n_1$ passes through 0, where the effect of sublayer polarized van Hove singularities should be most prominent (green curves in Fig.~\ref{SI_Fig4}). As in Fig.~5 from the main text, these cross-sections pass through the touching points of the tunneling region boundaries. Insets in Fig.~\ref{SI_Fig4} show the calculated tunneling conductance along these cross-sections, where the asymmetric tunneling resonance can be expected. As seen, the asymmetric resonance develops at twist angles $\theta\geqslant0.3^\circ$, because at smaller angles the touching point of the tunneling regions is located too close to the origin where fine features of the conductance map are smeared.

\bibliography{SI}